\documentclass[aps,
		prd,
		reprint,
        superscriptaddress,
        shortbibliography,
		  nofootinbib,
		floatfix,
		notitlepage, onecolumn
		]{revtex4-1}
\pdfoutput=1

\usepackage{
amsmath,
amssymb,
mathtools,
setspace,
IEEEtrantools,
tikz-feynman,
graphicx,
caption,
subcaption,
siunitx,
quantikz,
slashed,
comment
}

\usepackage[colorlinks=True, linkcolor=black, urlcolor=blue, citecolor=blue]{hyperref}
\graphicspath{ {./Images/} }

\renewcommand{\ket}[1]{\lvert #1 \rangle}
\renewcommand{\bra}[1]{\langle #1 \rvert}

\newcommand{\ketbra}[2]{\lvert #1 \rangle \langle #2 \rvert}
\renewcommand{\bar}{\overline}
\DeclareMathOperator{\arctanh}{arctanh}
\DeclareMathOperator{\atan}{atan2}

\newcommand{\volkovproparrow}{double, double distance=2pt, with arrow=0, arrow size=2.5pt}
\newcommand{\volkovprop}{double, double distance=2pt}

\def\XXint#1#2#3{{\setbox0=\hbox{$#1{#2#3}{\int}$ }
\vcenter{\hbox{$#2#3$ }}\kern-.6\wd0}}

\def\Zint#1{\mathchoice
    {\ZZint\displaystyle\textstyle{#1}}%
    {\ZZZint\textstyle\scriptscriptstyle{#1}}%
    {}{}%
    \!\int}
\def\ZZint#1#2#3{{\setbox0=\hbox{$#1{#2#3}{\int}$}
    \raise1.15ex\hbox{$#2#3$}\kern-.57\wd0}}
\def\ZZZint#1#2#3{{\setbox0=\hbox{$#1{#2#3}{\int}$}
    \raise0.85ex\hbox{$#2#3$}\kern-.53\wd0}}
\def\highdashint{\Zint-}

\begin{document}

\title{Hamiltonian truncation and quantum simulation of strong-field QED beyond tree level}

\author{Patrick Draper}
\email{pdraper@illinois.edu}
\affiliation{Department of Physics, University of Illinois, Urbana, IL 61801}
\author{Luis Hidalgo}
\email{luisdh2@illinois.edu}
\affiliation{Department of Physics, University of Illinois, Urbana, IL 61801}
\author{Anton Ilderton}
\email{anton.ilderton@ed.ac.uk}
\affiliation{Higgs Centre, School of Physics and Astronomy, University of Edinburgh, EH9 3FD, UK}

\begin{abstract}
	Quantum electrodynamics in strong background fields provides an interesting class of problems for classical and quantum simulation. In this paper we formulate simulations of polarization (helicity) flip for a photon colliding with a high-intensity plane wave. Polarization flip is a one loop effect, which requires addressing new issues that do not arise in simulations of tree-level processes. Working in the momentum-space Fock basis, while convenient for the extraction of scattering amplitudes, requires tuning counterterms to cancel large cutoff effects. We compute analytic formulas for the counterterms at one loop. We then construct circuits for quantum simulations of the process, perform noiseless simulations on classical computers to assess discretization errors, and discuss resource estimates for future simulations on quantum hardware.
\end{abstract}

\maketitle
\onecolumngrid

\section{Introduction}
	
Strong electromagnetic fields access processes that are not otherwise easily observed in quantum electrodynamics (QED), including e.g.~non-perturbative pair-production~\cite{schwinger:1951:pair_production} and vacuum polarization effects~\cite{Toll:1952rq, Narozhny:1968, Heinzl:2006xc, King:2010nka}, as well as a host of classical and quantum nonlinearities which are accessible with current laser technology~\cite{Cole:2017zca, Poder:2017dpw, Los:2024ysw, sarri2025inputeuropeanstrategyparticle}. While many tools have been developed to study `strong field QED' it remains challenging to account analytically for realistic background fields, higher multiplicity scattering, and higher loop corrections, see~\cite{Gonoskov:2021hwf, fedotov:2023:sfqed_review} for recent reviews.

In light of this it is interesting to explore numerical tools. In the low energy regime of quantum chromodynamics (QCD), lattice Monte Carlo methods provide high precision first-principles access to equilibrium properties and few-particle scattering amplitudes~\cite{Hansen:2025fbj}. Unfortunately, such methods are inapplicable to highly nonequilibrium dynamics (see however~\cite{Hebenstreit:2013qxa}), which represents one class of interesting problems in QCD, but essentially the entire class of interesting problems in SFQED. 

In the case of QCD, there has been considerable effort in recent years to develop quantum simulation techniques -- methods for studying real-time dynamics on a future quantum computer~\cite{l8b6-h5s5, davoudi2025quantumcomputationhadronscattering, PhysRevD.100.034518, 1km8-3tc3, PhysRevD.109.114510, PhysRevD.110.034515, ciavarella2025efficienttruncationssunclattice, PhysRevD.111.074516, PhysRevD.103.094501, balaji2025quantumcircuitssu3lattice, jiang2025nonabeliandynamicscubeimproving, schuhmacher2025observationhadronscatteringlattice, halimeh2025quantumsimulationoutofequilibriumdynamics}. Quantum computers offer new opportunities and new challenges for studying strongly coupled QFTs~\cite{PRXQuantum.4.027001, PRXQuantum.5.037001, Davoudi:2025kxb}. Although at present no quantum computer has been used to perform a Hamiltonian simulation that unambiguously outperforms classical simulations, the pace of development is rapid, and simulation techniques have improved in parallel with technological advances. It is hoped that on a timescale less than one career they will converge, and quantum computing will join classical lattice QCD in the compendium of essential computational tools~\cite{than2024phasediagramquantumchromodynamics, Bashore:2025uwb, chawdhry2024quantumalgorithmssimulationqcd, chawdhry2025quantumsimulationscatteringamplitudes}. 

Real time quantum simulation is also a natural line of attack on SFQED physics. In~\cite{hidalgo:2024:sfqed_quantum}, we took the first steps toward developing quantum simulation techniques for tree-level SFQED, using light-front quantization in the momentum basis. This formulation shares kinship with the Tamm-Dancoff approach~\cite{Tamm:1945qv, Dancoff:1950ud} and with the more recent Hamiltonian truncation program~\cite{PhysRevD.110.096016, schmoll2023hamiltoniantruncationtensornetworks, 10.21468/SciPostPhys.13.2.011}, within light-front quantization~\cite{PhysRevD.88.065014} and adapted for digital quantum computers~\cite{PhysRevD.111.016013, PhysRevA.105.032418, PhysRevD.111.096001}. In this work we consider quantum simulations of SFQED at loop-level, where new issues of renormalization must be confronted, and the encoded Hilbert subspace must be greatly enlarged. We focus on the photon polarization (or helicity) flip process~ \cite{Toll:1952rq,Narozhny:1968, Heinzl:2006xc, Meuren:2013oya, dinu:2014:helicity_flip, PhysRevD.111.016005, Karbstein:2022uwf, king:2023:vacuum_birefringence, Macleod:2024jxl}. This is the microscopic process underlying vacuum birefringence, in which there is significant experimental interest~\cite{Shen2018, Ahmadiniaz:2024xob, Smid:2025suu, Rinderknecht:2025zgu}. The one-loop contribution to polarization flip is shown in Fig.~\ref{one_loop_diagram}. Our objective here is \emph{not} to give new phenomenological insights into this process. Rather, polarization flip offers a natural next step beyond~\cite{hidalgo:2024:sfqed_quantum} because this process is (i) particularly simple in the class of background field we consider, where symmetries tightly constrain other degrees of freedom and (ii) related to that which we previously investigated -- the imaginary part of the no-flip amplitude is, by the optical theorem, equal to the total probability of nonlinear Breit-Wheeler~\cite{Borysov:2022cwc}.

We must emphasize that it is not at all obvious whether it is better to build quantum simulations of SFQED by working in momentum space or on a real spatial lattice~\cite{PhysRevD.11.395}. Both have significant advantages and disadvantages. Momentum space has the advantage that it is easier to ``carve out'' the relevant Hilbert space and avoid wasting quantum resources. It is also the natural basis to use for computing observable scattering amplitudes. Disadvantages of momentum space include some nonlocality in the interactions and significant complications related to renormalization and symmetry. The real-space lattice approach essentially reverses most of these properties. In this work we focus exclusively on light-front quantization in momentum space, and develop discretizations, truncations, and qubit encodings in this framework.

Hamiltonian renormalization on the light-front presents stimulating complications. The convenient light-cone gauge fixing and the natural regulators to use on a computer are non-covariant. This was first noted in~\cite{Mustaki:1990im}, in light-front QED in the absence of background fields. Fortunately, a small number of counterterms is sufficient to restore the theory to the correct universality class.

One of our main results is the computation of $A^2$-counterterms in the presence of background fields. We compute one-loop amplitudes with hard momentum cutoffs in two scenarios: first, perturbatively, for general weak backgrounds, where we immediately recover the nonlocal counterterm of~\cite{Mustaki:1990im} for zero background, as well as corrections which are background-field-dependent. Second, we consider the particular case of a (strong) impulsive plane wave background, which is both simple to work with analytically and the field for which we perform our classical and quantum simulations. Here we obtain the modified counterterms exactly, i.e.~to all orders in the background.

This paper is organized as follows. In Sec.~\ref{one_loop_momentum_space_hamiltonian} we introduce the lightfront Hamiltonian relevant to one-loop vacuum polarization processes. In Sec.~\ref{old_fashioned_perturbation_theory} we calculate the one-loop amplitudes for polarization (no) flip, and identify unphysical terms which arise due to the necessity of simulations requiring non-covariant cutoffs. We then find and discuss the appropriate counterterms to be added to the Hamiltonian. In Sec.~\ref{classical_simulation} we perform a classical Hamiltonian simulation. In Sec.~\ref{quantum_simulation} we discuss the resources needed to simulate this process on quantum computers. We conclude in Sec.~\ref{conclusion}.

\subsection{Notation and Conventions}
	
In this paper, we use light-front coordinates~\cite{Brodsky:1997de, harindranath1998introductionlightfrontdynamicspedestrians, Heinzl:2000ht} throughout. Four-vectors take the form $X^\mu = (X^+,X^-,X^1,X^2) \equiv (X^+,X^-,X^\perp)$, where $X^\pm = X^0 \pm X^3$. Meanwhile, $X_\mu = (X_+,X_-,X_\perp) = (\tfrac{X^-}{2},\tfrac{X^+}{2},-X^\perp)$. If $x^\mu$ and $p^\mu$ are coordinate and momentum four-vectors, then we take $x^+$ to be the light-front time coordinate and $p^-$ to be the light-front energy. We avoid the zero mode \cite{10.1143/PTP.56.270, heinzl2003lightconezeromodesrevisited}, and only use $p^\pm > 0$.

It is convenient to define three-vectors $\mathsf{x} = (x^-,x^1,x^2)$ and $\mathsf{p} = (p^+,p^1,p^2)$. Additionally, we will write $d^3\mathsf{x} = dx^-dx^1dx^2$ and $d^3\mathsf{p} = \theta(p^+) dp^+dp^1dp^2$. The on-shell dispersion relation for free particles is given by
\begin{equation}
    p^- = \frac{p^\perp p^\perp + m^2}{p^+} \;,
\end{equation}
in which repeated `$\perp$' indices are always summed over, and $p^-$ is the light-front energy. (Set $m=0$ for photons.)

Finally, we work in natural units with $\hbar=c=1$ and report values in $\si{\mega\electronvolt}$. $m=\SI{0.511}{\mega\electronvolt}$ is the mass of the electron and $e=0.303$ is chosen as the standard coupling value.
	
\section{One-Loop Momentum-Space Hamiltonian}
\label{one_loop_momentum_space_hamiltonian}

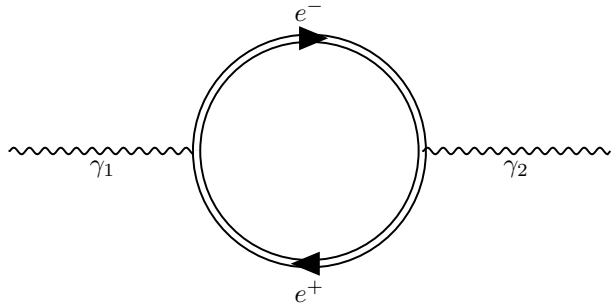
\begin{figure}[t]
    \begin{tikzpicture}[line width=0.75pt]
        \begin{feynman}
            \vertex (l);
            \vertex [right=2.5cm of l] (cl);
            \vertex [right=of cl] (c);
            \vertex [right=of c] (cr);
            \vertex [right=2.5cm of cr] (r);
            \vertex [above=of c] (ct);
            \vertex [below=of c] (cb);
            \vertex [above=1mm of ct] (ele) {$e^-$};
            \vertex [below=1mm of cb] (pos) {$e^+$};
            \diagram*{
                (l) -- [boson, edge label'=$\gamma_1$] (cl),
                (cl) -- [\volkovprop, quarter left] (ct) -- [\volkovproparrow, quarter left] (cr),
                (cr) -- [\volkovprop, quarter left] (cb) -- [\volkovproparrow, quarter left] (cl),
                (cr) -- [boson, edge label'=$\gamma_2$] (r)
            };
        \end{feynman}
    \end{tikzpicture}
    \caption{\textbf{One-loop Feynman diagram for polarization flip.} A photon with given polarization encounters a background field; unlike in vacuum, the photon can flip to an orthogonal polarization state through the one-loop interaction shown. Fermions interact directly with the background, and the double-line indicates that this interaction is accounted for non-perturbatively.}
    \label{one_loop_diagram}
\end{figure}
    
The SFQED Lagrangian density is
\begin{equation}
    \mathcal{L} = -\frac{1}{4} F_{\mu\nu} F^{\mu\nu} + \bar{\psi} ( i\slashed{\partial} - e\slashed{A} - e\slashed{\mathcal{A}} - m ) \psi \;,
\end{equation}
in which $\mathcal{A}^\mu$ is the background field potential. We will take this background to be a plane wave propagating in the $-\hat{z}$ direction. The wave may be parameterised by the two transverse components $\mathcal{A}^i(x^+)$, while $\mathcal{A}^\pm=0$.

The dependence of the background on light-front time $x^+$ makes it natural to work in light-front quantization. The corresponding Hamiltonian density $\mathcal{H}$ can be derived with the usual Legendre transform. The equations of motion yield constraints on the fermion and photon fields that can, unlike in `equal time' quantization, be solved explicitly and substituted back into $\mathcal{H}$~\cite{kogut:1970:infinite_momentum}. Essentially, by solving Gauss' law and fixing light-front gauge $A^+=0$, we end up with an entirely physical Hilbert space -- all states will satisfy Gauss' law. The resulting Hamiltonian depends on the dynamical fields $A^i$, i.e.~the transverse components of $A^\mu$, and on $\psi_-$, a projection of the original fermion field $\psi$ having only two nonzero components; explicitly $\psi_- = \tfrac14 \gamma^- \gamma^+ \psi$. The `price paid' for being able to work only with these unconstrained fields is that the Hamiltonian contains several new structures. Explicitly: 
\begin{IEEEeqnarray*}{rCl}
    H & = & \frac{1}{2} \int d^3\mathsf{x} \ \bigg(
    \frac{i}{2} \psi_-^\dag \frac{[ (\partial_i + ie\mathcal{A}_i)^2 - m^2 ]\psi_-}{\partial_-}
    + e \mathcal{A}^- \psi_-^\dag \psi_- 
    - \frac{1}{2} A^i \partial_\perp^2 A^i \\
    & + & \frac{e}{2} \partial_i A^i \frac{\psi_-^\dag \psi_-}{\partial_-} - \frac{e}{2} \psi_-^\dag \psi_- \frac{\partial_i A^i}{\partial_-} + \frac{iem}{2} \psi_-^\dag \gamma^i A_i \frac{\psi_-}{\partial_-} - \frac{iem}{2} \psi_-^\dag \gamma^i \frac{A_i \psi_-}{\partial_-} + \frac{e}{2} \psi_-^\dag \gamma^i \gamma^j \frac{\partial_i (A_j \psi_-)}{\partial_-} + \frac{e}{2} \psi_-^\dag \gamma^i \gamma^j A_i \frac{\partial_j \psi_-}{\partial_-}
    \\
     & + & \frac{ie^2}{2} \psi_-^\dag \gamma^i \gamma^j A_i \frac{A_j \psi_-}{\partial_-}  - \frac{e^2}{2} \psi_-^\dag \psi_- \frac{\psi_-^\dag \psi_-}{\partial_-^2}   \\
    & + & 
    \frac{ie^2}{2} \psi_-^\dag \gamma^i \gamma^j \mathcal{A}_i \frac{A_j \psi_-}{\partial_-}
    + \frac{ie^2}{2} \psi_-^\dag \gamma^i \gamma^j A_i \mathcal{A}_j \frac{\psi_-}{\partial_-} \bigg)\;. \yesnumber \label{H-full}
\end{IEEEeqnarray*}
The first line contains only `free' terms -- free in the sense that they are quadratic in the quantised fields, but describe non-trivial interactions between the fermions and the background $\mathcal{A}$. The second line contains three-point vertices, and are the same as in light-front QED without background. The third line contains the instantaneous (in light-front time) four-fermion interaction and the instantaneous two-photon-two-fermion interaction. They are consequences of solving the field constraint solutions, and again have the same form as without background. These will also be neglected below since they do not contribute to the processes of interest here~\cite{dinu:2014:helicity_flip}. The final line contains what are effectively three-point vertices in terms of the \emph{quantised} fields, corrected by the background. These are retained in what follows.   

We will work in a particular interaction picture -- the Furry picture~\cite{Furry:1951zz} -- in which the first line of (\ref{H-full}), containing the interaction of fermions with only the background, is treated exactly as a free Hamiltonian $H_0(x^+)$, while the remainder of (\ref{H-full}) constitutes the interaction Hamiltonian $V(x^+)$. The free part is, written out in terms of mode operators,
\begin{IEEEeqnarray*}{rCl}
    H_0(x^+) & = & \int \frac{d^3\mathsf{p}}{(2\pi)^3} \sum_{s=\pm\frac{1}{2}} \left[ \frac{(p^i-e\mathcal{A}^i(x^+))^2 + m^2}{p^+} \right] a_\mathsf{p}^{s\dag} a_\mathsf{p}^s +  \int \frac{d^3\mathsf{q}}{(2\pi)^3} \sum_{s=\pm\frac{1}{2}} \left[ \frac{(q^i+e\mathcal{A}^i(x^+))^2 + m^2}{q^+} \right] b_\mathsf{q}^{s\dag} b_\mathsf{q}^s \\
    & + & \int \frac{d^3\mathsf{k}}{(2\pi)^3} \sum_{\lambda=1,2} k^- \ c_\mathsf{k}^{\lambda\dag} c_\mathsf{k}^\lambda \;, \yesnumber
\end{IEEEeqnarray*}
in which repeated indices are always summed over, and the ladder operators $a_\mathsf{p}^{s\dag}/a_\mathsf{p}^s$, $b_\mathsf{q}^{s\dag}/b_\mathsf{q}^s$, and $c_\mathsf{k}^{\lambda\dag}/c_\mathsf{k}^\lambda$ describe electrons with helicity $-s$ and three-momentum $\mathsf{p}$, positrons with helicity $s$ and momentum $\mathsf{q}$, and photons with polarization $\lambda$ and momentum $\mathsf{k}$, respectively.

Turning to the interaction part, we neglect terms as described above, along with further three-point terms which do not contribute, namely the $e^\pm \leftrightarrow e^\pm + \gamma$ (photon emission / absorption) terms. What remains is
\begin{IEEEeqnarray*}{rCl}
    V(x^+) & = & e \int \frac{d^3\mathsf{k}}{(2\pi)^3 \sqrt{k^+}} \int \frac{d^3\mathsf{p}}{(2\pi)^3} \sum_{\lambda=1,2} \sum_{s=\pm\frac{1}{2}} \\
    & & \left\{ -m \epsilon^j_\lambda R^{-s,j} \left( \frac{1}{p^+} + \frac{1}{q^+} \right) a^{s\dag}_\mathsf{p} b^{s\dag}_\mathsf{q} c^\lambda_\mathsf{k} + m \bar{\epsilon}^i_\lambda R^{s,i} \left( \frac{1}{p^+} + \frac{1}{q^+} \right) c^{\lambda\dag}_\mathsf{k} b^s_\mathsf{q} a^s_\mathsf{p} \right. \\
    & + & \epsilon^j_\lambda \left[ \frac{p^{r'} - e\mathcal{A}^{r'}}{p^+} G^{-s,r',j} + \frac{q^{r'} + e\mathcal{A}^{r'}}{q^+} G^{-s,j,r'} + \frac{2k^j}{k^+} \right] a^{-s\dag}_\mathsf{p} b^{s\dag}_\mathsf{q} c^\lambda_\mathsf{k} \\
    & + & \left. \bar{\epsilon}^i_\lambda \left[ \frac{p^r - e\mathcal{A}^r}{p^+} G^{s,i,r} + \frac{q^r + e\mathcal{A}^r}{q^+} G^{s,r,i} + \frac{2k^i}{k^+} \right] c^{\lambda\dag}_\mathsf{k} b^{-s}_\mathsf{q} a^s_\mathsf{p} \right\} \yesnumber
    \label{interaction_hamiltonian}
\end{IEEEeqnarray*}
in which, from here on, momentum conservation fixes $\mathsf{q} = \mathsf{k}- \mathsf{p}$ for each Hamiltonian operator. $R^{s,i}$ and $G^{s,i,j}$ take the values $\pm 1$ or $\pm i$ only; working in the chiral basis of gamma matrices they are computed by
\begin{equation}
    R^{s,i} = w^{-s\dag} \gamma^i w^s \;,
    \qquad
    G^{s,i,j} = w^{s\dag} \gamma^i \gamma^j w^s \;,
\end{equation}
in which the $w$ are a spinor basis, $w^{(+\frac{1}{2})}=(0,1,0,0)$, $w^{(-\frac{1}{2})}=(0,0,1,0)$. Finally, $\epsilon^i_\lambda$ are a linear basis for the (transverse) polarization of the photon,
\begin{equation}
    \epsilon^i_1 = (1,0) \qquad \epsilon^i_2 = (0,1)
\end{equation}
throughout. The creation and annihilation operators obey the usual (anti-) commutation relations
\begin{equation}
    \{ a^s_\mathsf{p}, a^{s'\dag}_{\mathsf{p}'} \} = (2\pi)^3 \delta_{ss'} \delta^3(\mathsf{p}-\mathsf{p}') \qquad \{ b^s_\mathsf{q}, b^{s'\dag}_{\mathsf{q}'} \} = (2\pi)^3 \delta_{ss'} \delta^3(\mathsf{q}-\mathsf{q}') \qquad [ c^\lambda_\mathsf{k}, c^{\lambda'\dag}_{\mathsf{k}'} ] = (2\pi)^3 \delta_{\lambda\lambda'} \delta^3(\mathsf{k}-\mathsf{k}') \;,
\end{equation}
and create Fock states of particles.

\section{Amplitudes, renormalisation and counterterms}
\label{old_fashioned_perturbation_theory}

In this section we compute the one-loop helicity flip and no-flip amplitudes for a photon, momentum $k_\mu$, colliding with a general plane wave. We assume a head-on collision, taking $k_\mu = (k^+,0,0,0)$. We use old-fashioned perturbation theory in the Furry (interaction) picture for this calculation, since this echoes the form of the numerical calculations performed later. The interaction $V(x^+)$ thus becomes the Furry-picture interaction Hamiltonian $H_V(x^+)$; the transformation from $V$ to $H_V$ is given by the operator replacement $[ab] \to [ab] e^{-i\Phi}$, with `Volkov' phase~\cite{Wolkow:1935zz}
\begin{equation}
    \Phi(p,x^+) = \frac{1}{2} \left( \frac{1}{p^+} + \frac{1}{q^+} \right) \int_{0}^{x^+} dz^+ \left[ (p^1 - e\mathcal{A}^1(z^+))^2 + (p^2 - e\mathcal{A}^2(z^+))^2 + m^2 \right].
    \label{Volkov-phase}
\end{equation}
To compute a second-order contribution to a scattering amplitude we require the time evolution operator 
\begin{equation}
    U_V(+\infty,-\infty) = 1 - \frac{i}{2} \int_{-\infty}^{+\infty} dy^+ \ H_V(y^+) - \frac{1}{4} \int_{-\infty}^{+\infty} dy_1^+ \int_{-\infty}^{y_1^+} dy_2^+ \ H_V(y_1^+) H_V(y_2^+) + \cdots.
\end{equation}
The relation between $U_V$ and the Schr\"{o}dinger-picture time evolution operator, $U$, is $U_V = U_0^\dag U$, where $U_0$ is a time evolution operator derived with the free Hamiltonian $H_0$. Because $H_0$ is entirely diagonal in the Fock basis, $U_0$  simply applies a phase to Fock final states, which drops out of our probabilities. 

The polarization flip probability is 
\begin{equation}
    | \bra{\prescript{(2)}{k}{\gamma}} \hat{U}_V(+\infty,-\infty) \ket{\gamma_k^{(1)}} |^2
\end{equation}
The relevant part of the time evolution operator for $\gamma_{k}^{(1,2)}\rightarrow \gamma_{k}^{(1,2)}$ is
\begin{equation}
    \hat{U}_V(+\infty,-\infty) = 1 - \frac{e^2}{4} \int \frac{d^3\mathsf{k}}{(2\pi)^3 k^+} \sum_{\lambda,\lambda'=1,2} \bar{\epsilon}_{\lambda'}^i \Pi^{ij} \epsilon_\lambda^j \ c^{\lambda'\dag}_\mathsf{k} c^\lambda_\mathsf{k}
\end{equation}
where the on-shell polarization tensor $\Pi^{ij}$ has the explicit form~\cite{dinu:2014:helicity_flip}
\begin{IEEEeqnarray*}{rCl}
    \hspace{-1in} \Pi^{ij} & = & 2 \highdashint\!
    \frac{d^3\mathsf{p}}{(2\pi)^3}
    \int_{-\infty}^{+\infty} dy_1^+ \int_{-\infty}^{y_1^+} dy_2^+ \ e^{-i\Phi(p,y_1^+)} e^{i\Phi(p,y_2^+)} \yesnumber \label{piij} \\
    & \times & \left\{ m^2 \left( \frac{1}{p^+} + \frac{1}{q^+} \right)^2 \delta_{ij} + \left( \frac{1}{p^+} + \frac{1}{q^+} \right)^2 (p^r - e\mathcal{A}^r(y_1^+)) (p^r - e\mathcal{A}^r(y_2^+)) \delta_{ij} \right. \\
    & + & \left. \left( \frac{1}{p^+} - \frac{1}{q^+} \right)^2 (p^i - e\mathcal{A}^i(y_1^+)) (p^j - e\mathcal{A}^j(y_2^+)) - \left( \frac{1}{p^+} + \frac{1}{q^+} \right)^2 (p^j - e\mathcal{A}^j(y_1^+)) (p^i - e\mathcal{A}^i(y_2^+)) \right\} \;,
\end{IEEEeqnarray*}
in which, here and below, the `cut' integral sign denotes the presence of a factor $\theta(k^+-p^+)$ in the integrand. Note that $i \neq j$ ($i=j$) corresponds to the linear-polarization flip (no-flip) amplitude.

We expect UV divergences at loop level. For the one-loop helicity (no-) flip amplitudes these are dealt with by (i) regularising the integrals and (ii) subtracting the free-field contributions (the latter carry all the divergences). This assumes a covariant momentum regularisation scheme. There are, here, two relevant caveats which we now discuss.

First, regularisation in simulations is achieved by essentially non-covariant means, i.e.~hard momentum cutoffs are imposed, resulting in the explicit breaking of spacetime symmetries and gauge structure. Second, simulations in which we evolve an initial photon through discrete time steps will allow for polarization flip (or not) at each step, in other words a simulation does not compute a single amplitude, but rather a state. We must therefore renormalise in such a way that unwanted UV and symmetry-breaking effects are removed from the state. In a Wilsonian effective action obtained by integrating out UV quantum fluctuations, symmetry-breaking effects become encoded in local effective operators (which are particularly problematic if they are marginal or relevant). To deal with this, we can add counterterms to the bare Hamiltonian and tune then to cancel the unwanted effects of breaking due to the cutoff. We will see below that such operators are indeed generated in our system, and we will discuss the counterterms in details. To begin, we examine the polarization tensor in a specific choice of plane wave background which admits the most explicit calculations, and for which we will later perform both classical and quantum simulations. 
	
\subsection{Delta Pulses}

For analytic purposes, it is convenient to work with delta pulses~\cite{Fedotov:2013uja, Ilderton:2019vot, Ilderton:2024ufp}, in which the electric and magnetic fields are (sequences of) delta functions in light-front time. They can be thought of as infinite-frequency limits of narrow-width pulses. We use the background potential
\begin{equation}
    e\mathcal{A}^1(x^+) = e\mathcal{A}^2(x^+) = m\xi (\theta(x^+) - \theta(x^+-T)) \;,
    \label{deltapulses}
\end{equation}
which corresponds to an initial pulse at $x^+=0$, followed by a second pulse of equal magnitude and opposite sign at $x^+=T$. This magnitude is characterised by the dimensionless and tuneable parameter $\xi$, which is proportional to the classical work done on an electron by the background; an electron of initial momentum $\{p^+,p^i\}$ encountering the first delta pulse would be kicked to momentum $\{P^+,P^i\}$ with
\begin{equation}
    P^i = p^i - m\xi \;, \qquad P^+ = p^+ \;,
\end{equation}
a notation we adopt from here on to compactify expressions.

$\Pi^{ij}$ contains only two tensor structures; $\delta^{ij}$ for the no-flip amplitude $\mathcal{N}$, and $\mathcal{E}^{ij} = \epsilon_1^{(i}\epsilon_2^{j)}$ for the flip amplitude $\mathcal{F}$ (without a factor of $\frac{1}{2}$ in the symmetrization). Thus we can write
\begin{equation}
    \Pi^{ij} = \delta^{ij} \mathcal{N} + \mathcal{E}^{ij} \mathcal{F} \;.
\end{equation}
The flip amplitude $\mathcal{F}$ has both physical and unphysical terms. The physical piece is~\cite{Ilderton:2024ufp}
\begin{equation}
  \mathcal{F} \supset -8
  \highdashint \frac{d^3\mathsf{p}}{(2\pi)^3}
    \frac{4 q^+}{k^+}\left[ \frac{P^1}{P^-} - \frac{p^1}{p^-} \right]
    \left[ \frac{P^2}{P^-} - \frac{p^2}{p^-} \right] \left[ \frac{1 - e^{-i k^+ P^- T / (2q^+)}}{p^+k^+} \right] \;.
    \label{pi_ineqj_delta}
\end{equation}
This is the expression we use, below, to infer ideal Hamiltonian simulation parameters. The unphysical contribution is
\begin{equation}
    \mathcal{F} \supset  \frac{16iT}{k^+}
    \highdashint \frac{d^3\mathsf{p}}{(2\pi)^3}
    \frac{P^1 P^2}{P^\perp P^\perp+ m^2} \;.
    \label{linear_in_T_delta}
\end{equation}
It is straightforward to see that this contribution depends on how the integral is regulated. It can be cancelled by a counterterm which, we note, will depend on $\xi$. A third, seemingly divergent, contribution vanishes due to the residual symmetries of the cutoffs:
\begin{equation}
    \mathcal{F}  \supset
    \left( \int_{-\infty}^{0} dy_1^+ + \int_{T}^{+\infty} dy_1^+ \right)\frac{16i}{k
    ^+} \highdashint \frac{d^3\mathsf{p}}{(2\pi)^3} \frac{p^1 p^2}{p^\perp p^\perp+m^2}   = 0 \;.
\end{equation}
Turning to the no-flip amplitude (and again using residual symmetries to simplify expressions), we identify two contributions exhibiting unphysical cutoff effects:
\begin{IEEEeqnarray}{rCl}
    \mathcal{N} & \supset & -4i
    \left( \int_{-\infty}^{+\infty} dy_1^+ \right)
    \highdashint \frac{d^3\mathsf{p}}{(2\pi)^3} \left[ \frac{k^+}{p^+q^+}
    -
    \frac{2}{k^+}\frac{p^\perp p^\perp}{p^\perp p^\perp + m^2} \right] \label{divergence_delta} \\ [12pt]
    \mathcal{N} & \supset &
    \frac{8iT}{k^+}
    \highdashint \! \frac{d^3\mathsf{p}}{(2\pi)^3} \left[ \frac{P^\perp P^\perp}{P^\perp P^\perp +m^2} - \frac{p^\perp p^\perp}{p^\perp p^\perp +m^2} \right] .\label{noflip_delta}
\end{IEEEeqnarray}
Eq.~\eqref{divergence_delta} is independent of both $\xi$ and $T$, while \eqref{noflip_delta} has an overall factor of $T$, similar to \eqref{linear_in_T_delta}. Observe that both (\ref{linear_in_T_delta}) and (\ref{noflip_delta}) vanish if the background is turned off ($\xi\to0$), and would also vanish at non-zero $\xi$ if the integrals could be regulated to preserve transverse translation invariance. We write down the counterterms required to remove unwanted effects below. 

\subsubsection{Counterterms}
\label{counterterms}

The unphysical effects identified above break spacetime symmetries and violate the constraint algebra of QED in the Hamiltonian formalism. We discuss these violations in more detail in appendix~\ref{app:hamiltonian_field_theory}, but here we focus on removing them via the addition of counterterms. The counterterms are be associated with operators of the form $AA$, with a background-dependent coefficient function. We therefore define a counterterm Hamiltonian $H^c$ for the delta-pulse background to be
\begin{equation}
    H^c = \frac{1}{2} \int d^3\mathsf{x} \  C_{ij}[{\cal A}] A^i A^j \;,
\end{equation}
where
\begin{equation}
    \frac{1}{2} \int d^3\mathsf{x} \ A^i A^j = \int \frac{d^3\mathsf{k}}{(2\pi)^3 k^+} \sum_{\lambda,\lambda'=1,2} \left( \epsilon_{\lambda'}^i \bar{\epsilon}_\lambda^j \ c^{\lambda\dag}_\mathsf{k} c^{\lambda'}_\mathsf{k} + \bar{\epsilon}_{\lambda'}^i \epsilon_\lambda^j \ c^{\lambda'\dag}_\mathsf{k} c^\lambda_\mathsf{k} \right).
\end{equation}
$C_{ij}$ is chosen to cancel the unphysical contributions detailed above. We read from \eqref{divergence_delta}, \eqref{noflip_delta} and \eqref{linear_in_T_delta} that
\begin{equation}
\begin{split}
    C_{ij}[{\cal A}] &= \delta_{ij} e^2
    \highdashint  \frac{d^3\mathsf{p}}{(2\pi)^3} \left[ \frac{k^+}{p^+q^+} - \frac{2}{k^+}\frac{p^\perp p^\perp}{p^\perp p^\perp + m^2} \right] \\
    & -\delta_{ij} \Theta \frac{2e^2}{k^+}
    \highdashint 
    \frac{d^3\mathsf{p}}{(2\pi)^3} \left[ \frac{P^\perp P^\perp}{P^\perp P^\perp+m^2} - \frac{p^\perp p^\perp}{p^\perp p^\perp+m^2} \right]
    -\mathcal{E}_{ij} \Theta \frac{4e^2}{k^+}
    \highdashint
    \frac{d^3\mathsf{p}}{(2\pi)^3} \frac{P^1 P^2}{P^\perp P^\perp + m^2} \;,
\end{split}
\label{counterterms-new}
\end{equation}
where $\Theta$ is shorthand for the difference of step functions in the potential (\ref{deltapulses}). The first line of (\ref{counterterms-new}) is the same non-local counterterm as found in~\cite{Mustaki:1990im} for QED without background, and quantised on the lightfront (which serves as a minor consistency check on our calculations so far). The second line contains the field-dependent terms which only contribute between the delta pulses, and thus recover the the factors of $T$ in \eqref{noflip_delta} and \eqref{linear_in_T_delta}. Note that factors of $\frac{1}{k^+}$ are ultimately cancelled by the $p^+$ integral.

For simulations it is practical to evaluate the above integrals (or Riemann sums) numerically for finite $\Lambda$. However, it is also possible to compute them analytically in powers of $\frac{m}{\Lambda}$, which we record here. We find
\begin{equation}
    \frac{1}{2} \int_{-\Lambda}^{\Lambda} d^2p^\perp \frac{p^\perp p^\perp}{p^\perp p^\perp + m^2} \approx 2\Lambda^2 + m^2 \left( 2G + \pi \ln\left( \frac{m}{2\Lambda} \right) \right)
\end{equation}
which contributes to the field-independent term, and 
\begin{IEEEeqnarray*}{rCl}
    \int_{-\Lambda}^{\Lambda} d^2p^\perp \frac{P^1 P^2}{P^\perp P^\perp + m^2} & \approx & 2(m\xi)^2 - \frac{2(m\xi)^4 + 3m^2(m\xi)^2}{3\Lambda^2} \\ [12pt]
    \frac{1}{2} \int_{-\Lambda}^{\Lambda} d^2p^\perp \left[ \frac{P^\perp P^\perp}{P^\perp P^\perp +m^2} - \frac{p^\perp p^\perp }{p^\perp p^\perp +m^2} \right] & \approx & \frac{m^2(\pi+2)}{2\Lambda^2}(m\xi)^2 \yesnumber
    \label{mathematica_approximations}
\end{IEEEeqnarray*}
where $G$ is Catalan's constant.

\subsection{Effective Field Theory}

The unphysical contributions to the polarization tensor obtained above, and their counterterms, were computed for the specific background (\ref{deltapulses}). The breaking of symmetries by cutoffs is not, however, particular to this background. Before moving on to numerical simulations using delta pulses, we therefore briefly investigate the form of counterterms required in more general cases, namely general plane wave fields.

Observe first that in the delta-pulse case above, the counterterms are all-orders in $\xi$, but their field-dependent parts have perturbative expansions beginning at $\mathcal{O}(\xi^2)$. This, combined with the theta-functions present in (\ref{counterterms-new}), implies that to leading order in a weak-field expansion, the field-dependent counterterms originate from effective operators of the form $AA\mathcal{AA}$. We thefore proceed with an effective field theory approximation, expanding in \emph{weak fields and slow variations}. Clearly this second assumption excludes the delta-pulse case above, so we should not expect to find exact agreement with (the weak field limit of) the counterterms above, but we will see similar structures appearing.

For convenience define
\begin{equation}
    B_1^\ell(x^+) = \int_{0}^{x^+} dz^+ \ e\mathcal{A}^\ell(z^+) \qquad B_2(x^+) = \int_{0}^{x^+} dz^+ \ e^2 \mathcal{A}^\ell(z^+) \mathcal{A}^\ell(z^+)
\end{equation}
so that the phase factor (\ref{Volkov-phase}) may be written
\begin{equation}
    \Phi(p,x^+) = \frac{p^-+q^-}{2}x^+ - \left( \frac{1}{p^+} + \frac{1}{q^+} \right) p^\ell B_1^\ell(x^+) + \frac{1}{2} \left( \frac{1}{p^+} + \frac{1}{q^+} \right) B_2(x^+) \;.
\end{equation}
With this, we expand the product of phases $e^{-i\Phi(p,y_1^+)} e^{i\Phi(p,y_2^+)}$ appearing in the polarization tensor~\eqref{piij} up to quadratic order in $e\mathcal{A}^i$. This leaves a phase of $e^{i\frac{p^-+q^-}{2}(y_2^+-y_1^+)}$. Integrals containing this phase will of course be dominated by regions of slow oscillation, to identify which we note that
\begin{equation}
    \frac{p^-+q^-}{2}(y_2^+-y_1^+) \lesssim 1 \iff (y_2^+ - y_1^+) m \lesssim 2\left[ \frac{p^+q^+}{(k^+)^2} \right] \left[ \frac{k^+ m}{p^\perp p^\perp + m^2} \right] \;.
\end{equation}
The first factor in square brackets is $\leq \frac{1}{4}$. If we assume $k^+ \sim m$, then the second square-bracketed factor is $\lesssim 1$. It follows that for all loop momenta the phase oscillates rapidly unless $(y_2^+ - y_1^+) < \frac{1}{m}$, which justifies a light-front time-derivative expansion of background fields under the time integrals, as long as the background fields are slowly varying compared to the Compton wavelength. Near the extremes $p^+ \to 0$ and $p^+ \to k^+$, the approximations are even better, since then some energy is large and so $y_2^+$ and $y_1^+$ must be more localized. For effects that are generated by transverse momenta of order $\Lambda$,  $y_2^+-y_1^+$ is also more localized.

Thus we proceed by Taylor expanding about $y_2^+ = y_1^+$,
\begin{equation}
    B_1^\ell(y_2^+) \approx B_1^\ell(y_1^+) + e\mathcal{A}^\ell(y_1^+) y_m^+ \qquad B_2(y_2^+) \approx B_2^\ell(y_1^+) + e^2\mathcal{A}^\ell(y_1^+) \mathcal{A}^\ell(y_1^+)y_m^+
\end{equation}
where we henceforth use a change of variables: $y_p^+ = y_2^+ - y_1^+$ and $y_m^+ = y_2^+ - y_1^+$. Neglecting derivatives of the background field, we will approximate $e\mathcal{A}^i(y_2^+) \approx e\mathcal{A}^i(y_1^+)$.

With this procedure, we obtain
\begin{equation}
    \Pi^{ij}_{(i \neq j)} = \frac{32i}{(2\pi)^3} \frac{1}{2+\frac{m^2}{\Lambda^2}} \int_{-\infty}^{+\infty} dy_p^+ \ e^2\mathcal{A}^i(\tfrac{y_p^+}{2}) \mathcal{A}^j(\tfrac{y_p^+}{2})
\end{equation}
for $i \neq j$ and
\begin{IEEEeqnarray*}{rCl}
    \Pi^{ij}_{(i=j)} & = & -2i \delta_{ij}
    \highdashint 
    \frac{d^3\mathsf{p}}{(2\pi)^3} \int_{-\infty}^{+\infty} dy_p^+ \left\{ \frac{k^+}{p^+q^+} - \frac{2}{k^+} \frac{p^\perp p^\perp}{p^\perp p^\perp + m^2} \right\} \\
    & + & \frac{16i}{(2\pi)^3} \int_{-\infty}^{+\infty} dy_p^+ \ e^2\mathcal{A}^i(\tfrac{y_p^+}{2}) \mathcal{A}^i(\tfrac{y_p^+}{2}) \left\{ \frac{1+2\frac{m^2}{\Lambda^2}}{\sqrt{(1+\frac{m^2}{\Lambda^2})^3}} \arctan\left( \tfrac{1}{\sqrt{1+\frac{m^2}{\Lambda^2}}} \right) - \frac{1}{1 + \frac{m^2}{\Lambda^2}} + \frac{1}{2 + \frac{m^2}{\Lambda^2}} \right\} \\
    & + & \frac{16i}{(2\pi)^3} \int_{-\infty}^{+\infty} dy_p^+ \ e^2\mathcal{A}^{3-i}(\tfrac{y_p^+}{2}) \mathcal{A}^{3-i}(\tfrac{y_p^+}{2}) \left\{ \frac{1}{2+\frac{m^2}{\Lambda^2}} - \frac{1}{\sqrt{1+\frac{m^2}{\Lambda^2}}} \arctan\left( \tfrac{1}{\sqrt{1+\frac{m^2}{\Lambda^2}}} \right) \right\} \yesnumber
\end{IEEEeqnarray*}
for $i=j$, to all orders in $\frac{m}{\Lambda}$. (Note the use of $y_m^+ \to 0$, hence $y_1^+ \to \frac{y_p^+}{2}$.) Expanding these amplitudes to $\mathcal{O}(\frac{m^2}{\Lambda^2})$ yields expressions analogous to those in \eqref{mathematica_approximations}. 

As in the delta pulse analysis, we observe in these expressions UV-finite (independent of $\Lambda$ as $\Lambda\rightarrow\infty$) contributions associated with verboten effective operators of the form $AA$ and $AA\mathcal{AA}$. This phenomenon appears to be a general deficiency of the hard momentum cutoffs that are most natural for numerical analysis. In contrast physical effects are associated with higher-order effective operators, in particular terms with more derivatives, and they can be derived by keeping more powers of $e\mathcal{A}^i$ and its derivatives in the Taylor expansions. Here we capture only the counterterms needed to remove the unphysical $AA$ and $AA\mathcal{AA}$ contributions.

We define a generalized counterterm Hamiltonian:
\begin{equation}
    H^c = \frac{1}{4} \int d^3\mathsf{x} \ \bigg( C_{AA} A_\mu A^\mu + C_{AA\mathcal{AA}} (A_\mu A^\mu) (\mathcal{A}_\nu \mathcal{A}^\nu) + C_{2A\mathcal{A}} (A_\mu \mathcal{A}^\mu)^2 + D_{\mu\nu\rho\sigma} A^\mu A^\nu \mathcal{A}^\rho \mathcal{A}^\sigma \bigg)
\end{equation}
After gauge-fixing, this becomes
\begin{equation}
    H^c = \frac{1}{4} \int d^3\mathsf{x} \ \bigg( -C_{AA} A^i A^i + C_{AA\mathcal{AA}} A^i A^i \mathcal{A}^j \mathcal{A}^j + C_{2A\mathcal{A}} A^iA^j\mathcal{A}^i\mathcal{A}^j + D_{ijk\ell} A^i A^j \mathcal{A}^k \mathcal{A}^\ell \bigg)
\end{equation}
Using our EFT calculations above, the counterterm coefficients are:
\begin{IEEEeqnarray*}{rCl}
    C_{AA} & = & -e^2
    \highdashint 
    \frac{d^3\mathsf{p}}{(2\pi)^3} \left[ \frac{k^+}{p^+q^+} - \frac{2}{k^+} \frac{p^\perp p^\perp}{p^\perp p^\perp + m^2} \right] \\ [12pt]
    C_{AA\mathcal{AA}} & = & -\frac{8e^4}{(2\pi)^3} \left[ \frac{1}{2+\frac{m^2}{\Lambda^2}} - \frac{1}{\sqrt{1+\frac{m^2}{\Lambda^2}}} \arctan\left( \tfrac{1}{\sqrt{1+\frac{m^2}{\Lambda^2}}} \right) \right] \\ [12pt]
    C_{2A\mathcal{A}} & = & -\frac{8e^4}{(2\pi)^3} \left[ \left( \frac{1+2\frac{m^2}{\Lambda^2}}{\sqrt{(1+\frac{m^2}{\Lambda^2})^3}} + \frac{1}{\sqrt{1+\frac{m^2}{\Lambda^2}}} \right) \arctan\left( \tfrac{1}{\sqrt{1+\frac{m^2}{\Lambda^2}}} \right) - \frac{1}{1 + \frac{m^2}{\Lambda^2}} \right] \\ [12pt]
    D_{ijk\ell} & = & - \mathcal{E}^{ij} \mathcal{E}^{k\ell} \left( \frac{8e^4}{(2\pi)^3} \frac{1}{2 + \frac{m^2}{\Lambda^2}} + C_{2A\mathcal{A}} \right) \yesnumber
\end{IEEEeqnarray*}

We remark that expansions analogous to those above lie behind the `locally constant field approximation', one of the central tools for linking SFQED predictions, simulations and experiments~\cite{Blackburn:2025vnv}. For a review see~\cite{Fedotov:2013uja} and for discussions of the failings and extension of the approximation see~\cite{DiPiazza:2017raw, Ilderton:2018nws}.

\section{Classical Simulation}
\label{classical_simulation}

We now have counterterms that we can include into our Hamiltonian to remove unwanted cutoff effects during time evolution. We have also shown that these are consistent with the expected form of counterterms in more general fields. Therefore, `real-time' classical and quantum simulations alike should be robust against cutoff effects, as long as the counterterms are present in the time evolution operator. In simulations, this operators may be approximated~\cite{J_Huyghebaert_1990} by the first-order time evolution Trotterization
\begin{equation}
    U(x^+,x_0^+) = \mathcal{T} e^{-\frac{i}{2} \int_{x_0^+}^{x^+} H(y^+) \ dy^+} = e^{-\frac{i}{2}H(x_n^+)\Delta x^+} \cdots e^{-\frac{i}{2}H(x_1^+)\Delta x^+} + \mathcal{O}[(\Delta x^+)^2]
    \label{time_evo_trotter}
\end{equation}
for $n$ time steps of size $\Delta x^+ = \frac{x^+-x_0^+}{n}$ and $n$ times $x_0^+ < x_1^+ < \cdots < x_n^+ = x^+$, with $x_k^+ = k\Delta x^+ + x_0^+$.

We wish to compute the probability $| \bra{\prescript{(2)}{k}{\gamma}} U(+\infty,-\infty) \ket{\gamma_k^{(1)}} |^2$ with Hamiltonian simulation~\cite{PhysRevD.88.065014, PhysRevD.102.016017, PhysRevD.95.096012} via Trotterization. To deal with the asymptotic limits, we perform an adiabatic time evolution. This splits the time-evolution of the system into three regions: adiabatic turn-on (from $X^+_{\text{on},i}$ to $X^+_{\text{on},f}$), constant coupling (from $X^+_{\text{on},f}$ to $X^+_{\text{off},i}$), and adiabatic turn-off (from $X^+_{\text{off},i}$ to $X^+_{\text{off},f}$).
	
During adiabatic turn-on we begin in a single-photon Fock state, which is an eigenstate of $H_0$, the free Hamiltonian. We turn on the coupling $e(x^+)$ linearly during this interval. This allows us to approximately prepare an eigenstate of $H$, the full Schr\"{o}dinger-picture Hamiltonian. During constant coupling, the interaction with the background field occurs. The coupling $e(x^+)$ is held constant. During adiabatic turn-off we linearly turn off the coupling $e(x^+)$. This returns the eigenbasis to Fock space. Thus, we can measure overlap with a final single-photon Fock state.

\subsection{Hamiltonian Truncation}

To simulate on a computer, we must work with a finite Hilbert space. We include a tower of two-particle $e^+e^-$ states, and \emph{two} one-particle photon states, with polarization $\lambda=1,2$ and as above, momentum $k^\mu = (k^+,0,0,0)$ corresponding to a head-on collision between the incident photon and our plane wave. The restriction to \emph{only} this momentum mode follows from the explicit results for the polarization and time-evolution operators above, and the symmetries of the plane wave background; while only three components of momentum are conserved, this is enough to fix the on-shell momentum of a single incoming photon. Internal degrees of freedom can still change due to interaction with the background, though, in particular the helicity of the photon.

Momentum is discretized, turning integrals into Riemann sums:
\begin{equation}
    \int\!\frac{d^3\mathsf{p}}{(2\pi)^3} F(p) [a] \to \sum_{n_+=1}^{+\infty} \sum_{n_1,n_2=-\infty}^{+\infty} \frac{1}{L_+ L_\perp^2} F(\tfrac{2\pi n}{L}) [\sqrt{L_+L_\perp^2} a].
\end{equation}
where $[a]$ denotes some single-particle operator that is scaled by $\sqrt{L_+L_\perp^2}$ when passing from the continuum to finite volume. The momentum-space lattice with longitudinal spacing $\frac{2\pi}{L_+}$ and transverse spacing $\frac{2\pi}{L_\perp}$ recovers the continuum when $L \to +\infty$. Electron momenta now appear as $\mathsf{p} = (\frac{2\pi n_+}{L_+}, \frac{2\pi n_1}{L_\perp}, \frac{2\pi n_2}{L_\perp})$. To truncate the Hamiltonian, we impose cutoffs on the momentum modes. In the longitudinal direction, we allow up to $K-1$ modes. In the transverse directions, we use symmetric cutoffs $\Lambda = \frac{2\pi N}{L_\perp}$ for $N \geq 0$:
\begin{equation}
    \sum_{n_+=1}^{+\infty} \sum_{n_1,n_2=-\infty}^{+\infty} \frac{1}{L_+ L_\perp^2} F(\tfrac{2\pi n}{L}) [\sqrt{L_+L_\perp^2} a] \to \sum_{n_+=1}^{K-1} \sum_{n_1,n_2=-N}^{N} \frac{1}{L_+ L_\perp^2} F(\tfrac{2\pi n}{L}) [\sqrt{L_+L_\perp^2} a].
\end{equation}
For a given $(K,N)$, the number of electron-positron pairs is $4(K-1)(2N+1)^2$, where the factor of four comes from counting spin/helicity.

We use the usual occupation number basis, and order the states as
\begin{equation}
    \ket{\dots, {e^-}^{(+\frac{1}{2})}_{\mathsf{p}_n}, \dots, {e^-}^{(-\frac{1}{2})}_{\mathsf{p}_n}, \dots, {e^+}^{(+\frac{1}{2})}_{\mathsf{q}_n}, \dots, {e^+}^{(-\frac{1}{2})}_{\mathsf{q}_n}, \dots, \gamma^{(1)}_{\mathsf{k}_n}, \dots, \gamma^{(2)}_{\mathsf{k}_n}, \dots} \;.
    \label{fock_state}
\end{equation}
Usually, a fermion creation and annihilation operator acts with a factor of $(-1)^\zeta$, where $\zeta$ is the sum of fermion occupation numbers to the left of the fermion acted on, relative to the ordering in \eqref{fock_state}. $[a^\dag a]$- and $[b^\dag b]$-type operators will always yield factors of $(-1)^{2\zeta}=1$. Meanwhile, $[a^\dag b^\dag]$- and $[ba]$-type operators that appear in our normal-ordered Hamiltonian will act only on vacuum or electron-positron pair Fock states (in our eventual truncation). The $[b^\dag]$ operator in $[a^\dag b^\dag]$ first acts on the vacuum, where $(-1)^\zeta \to 1$. Then $[a^\dag]$ acts on a positron Fock state, but due to the ordering in \eqref{fock_state}, positron occupation numbers do not contribute to $\zeta$ for $[a^\dag]$ operators. A similar situation occurs for $[ba]$. Therefore, the $(-1)^\zeta$ factor will always result in factors of one in our simulations and can be neglected.

\subsection{Simulation Parameters}

\begin{figure}[t]
    \centering
    \includegraphics[width=0.75\textwidth]{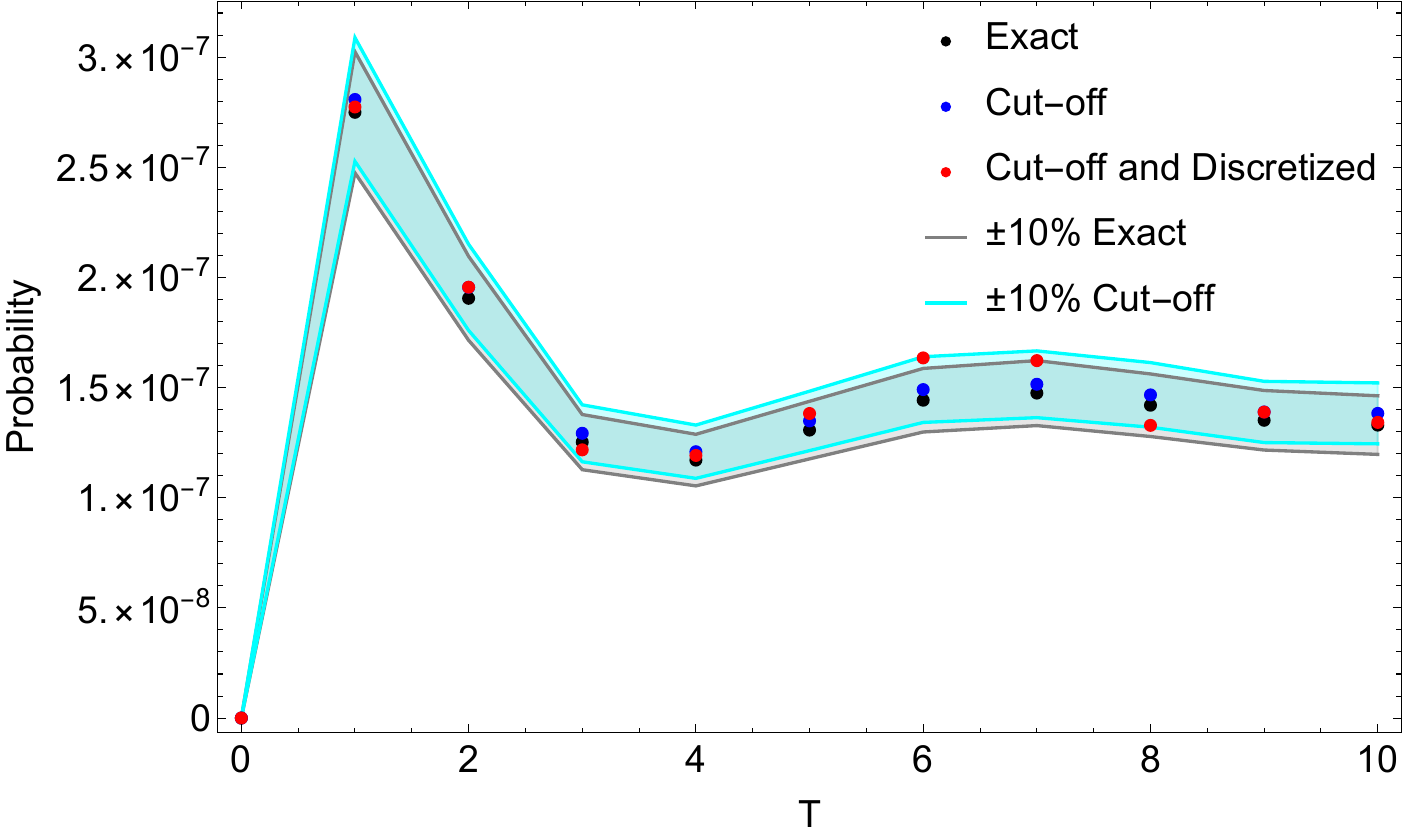}
    \caption{\textbf{Quantifying cutoff and discretization effects.} We compare the accuracy of the cut-off polarization flip probability, calculated with \eqref{cut_off_amp} and $\Lambda=5m$, with the exact probability from \eqref{pi_ineqj_delta}. We also compare the cut-off and discretized probability, found with \eqref{cut_off_disc_amp} and $(K,N)=(12,13)$, with the cut-off probability. Here, $(k^+,\xi)=(m,1)$ and the delta-pulse spacing $T$ is varied. The band marked ``$\pm 10\%$ Exact'' shows the window in which the exact probability varies by 10\%. The band marked ``$\pm 10\%$ Cut-off'' shows the same around the probability computed with the cut-off amplitude.}
    \label{param_analysis}
\end{figure}

Returning to the delta pulses, we seek appropriate parameters $\{ \Lambda,K,N \}$ given inputs $\{ T,k^+,\xi \}$. We choose $k^+=m$ and $\xi=1$, while $T \sim \mathcal{O}(\frac{1}{m})$. Note that choosing $(K,N)$ fixes $L_+ = \frac{2\pi K}{k^+}$ and $L_\perp = \frac{2\pi N}{\Lambda}$.
	
We can numerically integrate \eqref{pi_ineqj_delta} to get ``exact'' polarization flip probabilities. Then, by imposing our UV cutoff $\Lambda$, we calculate
\begin{equation}
    -8 \int_{0}^{k^+} \frac{dp^+}{2\pi} \int_{-\Lambda}^{\Lambda} \frac{d^2p^\perp}{(2\pi)^2} 
    \frac{4q^+}{k^+}\left[ \frac{P^1}{P^-} - \frac{p^1}{p^- } \right] \left[\frac{P^2}{P^-} - \frac{p^2}{p^-} \right] \left[ \frac{1 - e^{-i k^+ P^- T/(2q^+)}}{p^+k^+} \right]
    \label{cut_off_amp}
\end{equation}
to obtain ``cut-off'' probabilities. We find that $\Lambda=5m$ offers good agreement between exact and cut-off probabilities.
	
Next, we calculate the Riemann sum of \eqref{cut_off_amp} with $(K,N)$:
\begin{equation}
    -\frac{8}{L_+L_\perp^2} \sum_{n_+=1}^{K-1} \sum_{n_i,n_j=-N}^{N} \frac{4q^+}{k^+}\left[ \frac{P^1}{P^-} - \frac{p^1}{p^- } \right] \left[\frac{P^2}{P^-} - \frac{p^2}{p^-} \right] \left[ \frac{1 - e^{-i k^+ P^- T/(2q^+)}}{p^+k^+} \right]
    \Bigg|_{
    p^i \to \frac{2\pi n_i}{L_\perp} \;,\;
    p^+ \to \frac{2\pi n_+}{L_+}
    }
    \label{cut_off_disc_amp}
\end{equation}
to get ``cut-off and discretized'' probabilities. We find that $(K,N)=(12,13)$ offers fair agreement with the cut-off probabilities. The complete analysis is shown in figure~\ref{param_analysis}.

\subsection{Results}

\begin{figure}[t]
    \centering
    \includegraphics[width=\textwidth]{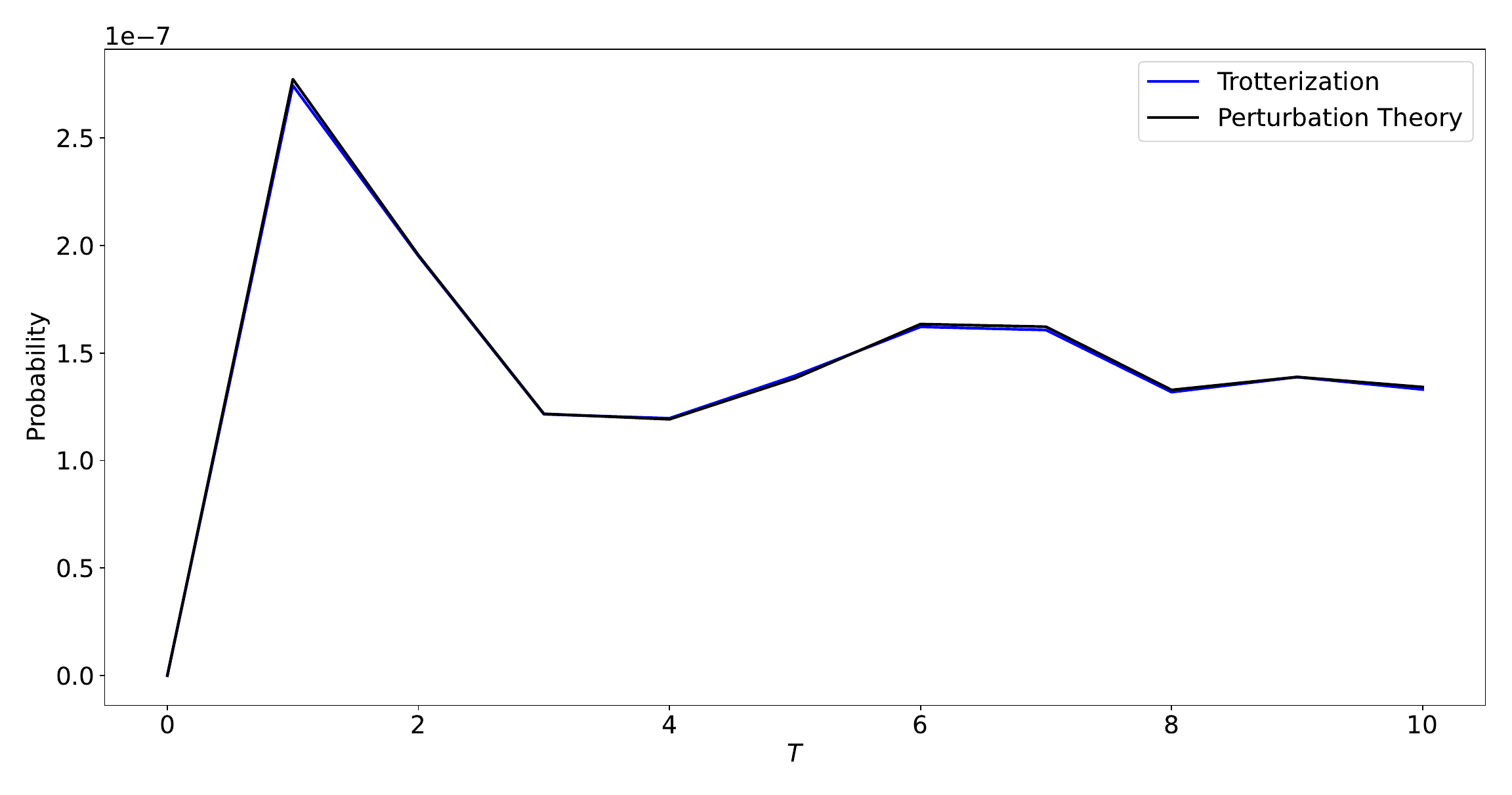}
    \caption{\textbf{Convergence of Trotter simulation and perturbation theory.} The figure above compares  polarization flip probabilities obtained from Trotterization versus cut-off and discretized perturbation theory for several values of $T$. Probabilities are obtained from the late-time state after a period of adiabatic turn-off of the coupling. See figure~\ref{classical_sim} for a sample simulation that produced the Trotterization data point at $T=5$.}
    \label{pert_and_trot}
\end{figure}

\begin{figure}[t]
    \centering
    \includegraphics[width=\textwidth]{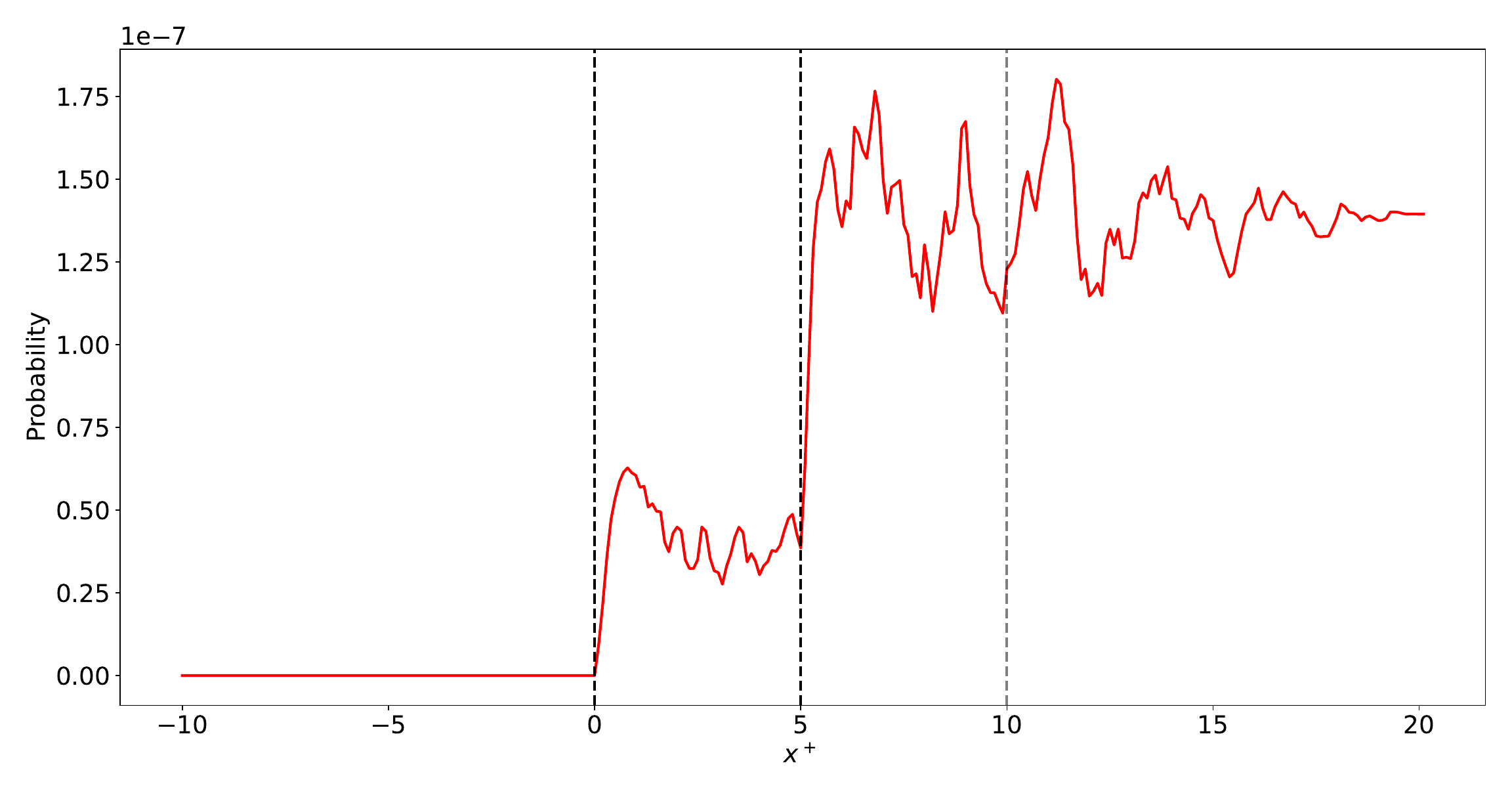}
    \caption{\textbf{Time-dependent polarization flip probabilities obtained from Trotterization.} Here, $\Delta x^+ = 0.1$ and $T=5$. Adiabatic turn-on starts at $x^+=-10$ and ends at $x^+=0$. From $x^+=0$ to $x^+=10$, the coupling is held at $e=0.303$. The black dashed lines show where an initial delta pulse appears at $x^+=0$ and is later canceled by an opposite-sign pulse at $x^+=5$. The gray dashed line shows when adiabatic turn-off begins, lasting until $x^+=20$ where $e=0$.}
    \label{classical_sim}
\end{figure}

All counterterms detailed in Sec.~\ref{counterterms} must be implemented -- even the no-flip ones. This is because Hamiltonian simulation will produce more dynamics than figure~\ref{one_loop_diagram} alone. For example, the simulation will allow processes such as $\gamma_1 \to \gamma_2 \to \gamma_1 \to \gamma_1 \to \cdots$. The counterterms are calculated numerically as Riemann sums. Figure~\ref{pert_and_trot} shows polarization flip probabilities for different values of $T$ using Trotterization versus a cut-off and discretized perturbative scattering amplitude. Meanwhile, figure~\ref{classical_sim} shows the underlying ``real-time'' polarization flip probability for $T=5$.

\begin{figure}[t]
    \centering
    \includegraphics[width=\textwidth]{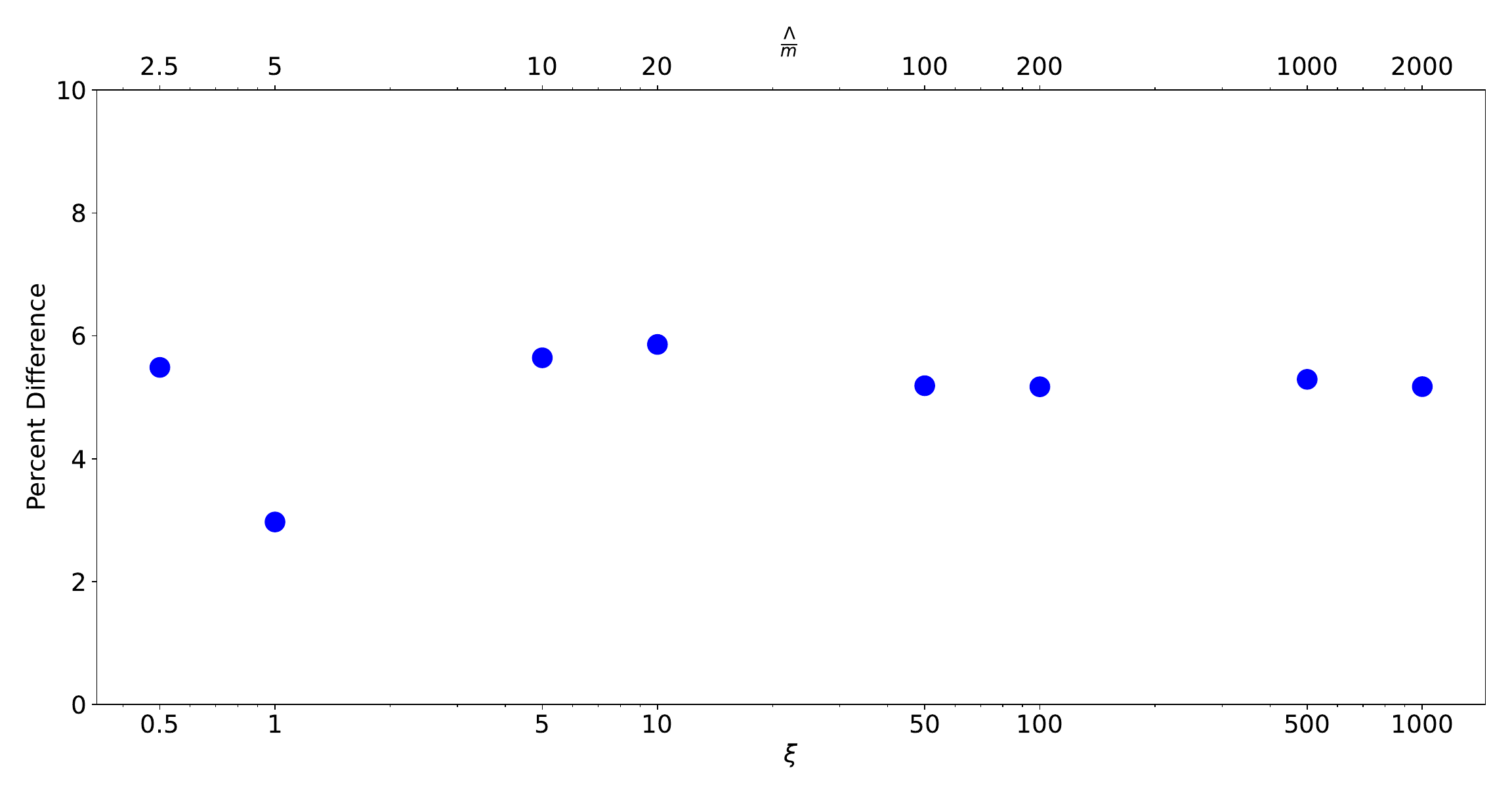}
    \caption{\textbf{Robustness of the discretization procedure to changes in background field strength.} This figure shows the percent difference of cut-off and discretized probabilities, averaged over a large-$T$ interval, with the expected probabilities calculated with \eqref{large_T} for various values of $\xi$. $(K,N)=(12,13)$ was kept fixed, but $\Lambda$ was adjusted for each $\xi$.}
    \label{many_xi}
\end{figure}

It is worth noting that the numerical results shown in figures~\ref{param_analysis} and \ref{pert_and_trot} agree well with the expected long-$T$ behavior. For large $T$, the phase in \eqref{pi_ineqj_delta} can be dropped, and the polarization flip probability becomes
\begin{equation}
    \lim\limits_{T\to+\infty} |_k^2 \bra{\gamma} \hat{U}_V(+\infty,-\infty) \ket{\gamma}_k^1|^2 = \frac{e^4}{36\pi^4 \xi^2} \left[ \xi - \frac{2}{\sqrt{\xi^2+2}} \arctanh\left( \frac{\xi}{\sqrt{\xi^2+2}} \right) \right]^2
    \label{large_T}
\end{equation}
With $\xi=1$, this evaluates to about $1.38 \times 10^{-7}$. In figure~\ref{many_xi}, we further demonstrate the robustness of our simulation parameters to variations in $\xi$.

\clearpage

\section{Quantum Simulation}
\label{quantum_simulation}

Quantum computers may one day be useful for simulating the large number of degrees of freedom found in a many-body quantum system. This is because their (typical) fundamental unit of information, the qubit, is itself a quantum system~\cite{Nielsen_Chuang_2010}. A qubit has two basis states: the $\ket{0} = \binom{1}{0}$ state and the $\ket{1} = \binom{0}{1}$ state, which are eigenstates of the Pauli $Z$ operator\footnote{In the quantum computing community it is common to denote the Pauli matrices $\sigma^1,\sigma^2,\sigma^3$ as $X,Y,Z$, respectively.}. When $n$ qubits are brought together, their $2^n$-dimensional Hilbert space can be described using the computational basis $\{ \ket{0 \dots 0}, \ket{0,\dots,1}, \dots, \ket{1 \dots 1} \}$, which is a basis of tensor-product states. Therefore, if the target quantum system to be simulated has a Hilbert space with dimension $d$, then at least $n = \log_2 \lceil d \rceil$ qubits are needed to encode the basis states of the target quantum system.

From here, most quantum computers work by applying norm-preserving unitary operators, called quantum gates, on qubits. (We direct the reader to~\cite{PhysRevA.52.3457} for a list of common quantum gates.) Note that most quantum gates act nontrivially on only a subset of the qubits; in fact, any quantum computation involving $n$ qubits can be performed with only single-qubit and two-qubit gates. The complete product of quantum gates used in a quantum computation is called the quantum circuit. In this section we detail how the truncated Hilbert space in the polarization flip process may be encoded with qubits, as well as how the quantum circuit that implements the time evolution operator may be built for our encoding.

The Hamiltonian as formulated in Sec.~\ref{one_loop_momentum_space_hamiltonian} acts on Fock space, a convenient basis of which is the occupation number basis. Basis states appear in the form $\ket{n_1 n_2 \cdots n_N}$ for an $N$-particle system, where $n_i$ labels the occupation number of the $i^\text{th}$ particle species. The total number of particles in a basis state is $\sum_i n_i$, in which $n_i=0,1$ for fermions and $n_i = 0,1,2,\dots$ for bosons. For our quantum simulations, we will truncate the maximum photon occupation number to one, as we did in Sec.~\ref{classical_simulation}. Therefore, all remaining basis states are tensor products of $N$ $\ket{0}$ and $\ket{1}$ states, which is exactly the form of a computational basis state for $n=N$ qubits. Thus, $n$ qubits can encode the Fock space for $n$ single-particle species, with each computational basis state corresponding to an occupation number basis state. Effectively, each qubit encodes the state of a specific particle species. For example, if $n=3$, then $\ket{000}$ would be the vacuum state, $\ket{001}$ could be a single-photon state, and $\ket{110}$ could be an electron-positron pair state. The photon creation and annihilation operators appearing in the Hamiltonian are simply ladder operators that flip qubit basis states $\ket{0} \leftrightarrow \ket{1}$ or annihilate them. The fermion creation and annihilation operators, on the other hand, must be formulated to act on qubits so as to maintain their anticommutation relations.

The Jordan-Wigner transformation \cite{PhysRevA.65.042323, seeley:2012:fermion_to_qubits, PhysRevB.109.115149} is a commonly used tool to map fermionic operators to qubit operators. It requires each single-particle fermion state to be encoded with its own qubit, where $\ket{0}$ means the fermion is absent and $\ket{1}$ means the fermion is present. Fermionic anticommutation relations are upheld by attaching a tensor-product of $Z$ operators in front of the ladder operator according to a prescribed ordering of the fermions, as in \eqref{fock_state}. For the one-loop polarization flip process, we find this to be an unfeasible approach. Thousands of intermediate electron-positron pairs would require thousands of qubits -- not currently available on quantum computers -- to simulate. In the future, when large-scale fault-tolerant quantum computers become available, this may be the encoding of the Fock space necessary to simulate more general SFQED processes such as cascades. For now, it is too expensive to carry out a proof-of-principle quantum simulation of our polarization flip process with it.

Instead, we recognize that our Hamiltonian truncation leaves us with only electron-positron and single-photon Fock basis states. We can encode a total of $d$ states with $n = \log_2\lceil d \rceil$ qubits, as described above. Then a computational basis state such as $\ket{0011}$, implying the use of four qubits, could encode (for example) some electron-positron pair when $8 < d \leq 16$. However, such a dense binary encoding can lead to a rather expensive quantum circuit. The controlled unitaries detailed below would generally have to be controlled on all $n$ qubits, which leads to more single- and two-qubit gates in their decomposition. In particular, more $T = \sqrt[4]{Z}$ gates -- deemed costly in fault-tolerant quantum computing for their difficulty to simulate classically -- would be required.

\begin{figure}[h]
    \centering
    \includegraphics[width=0.5\textwidth]{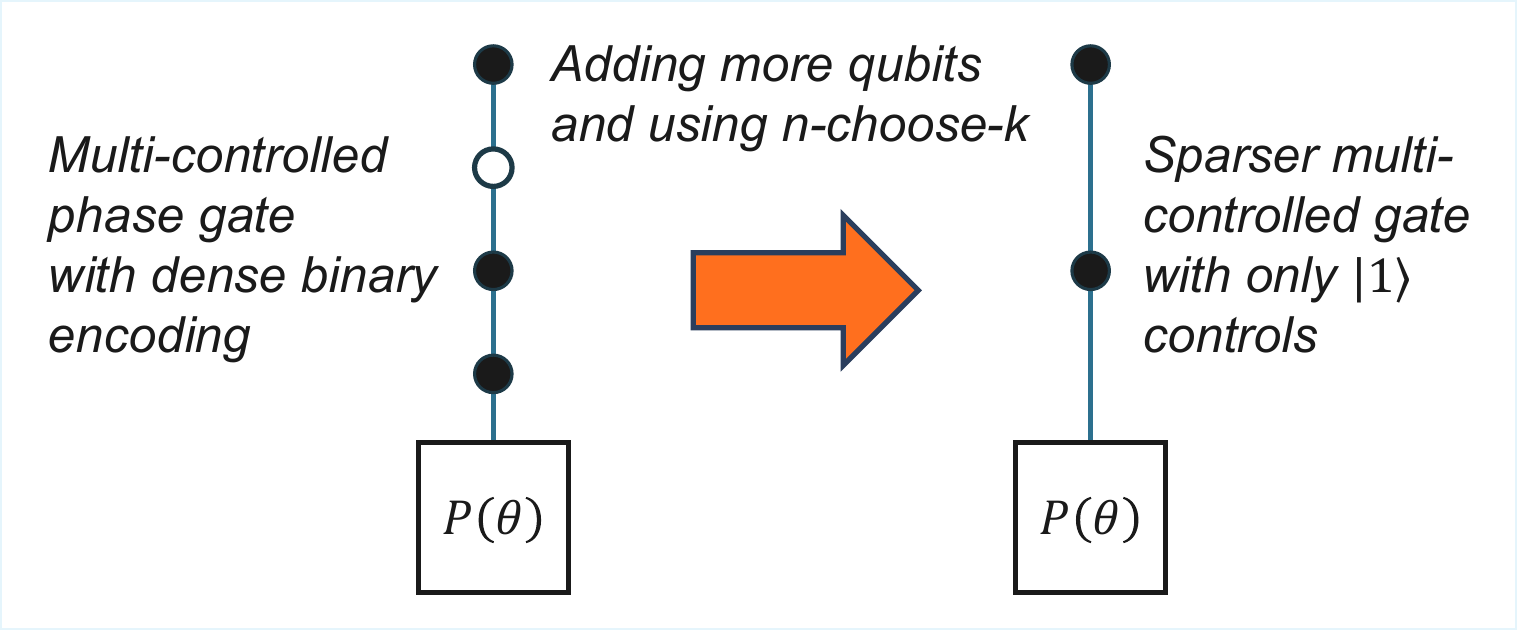}
    \caption{\textbf{Dense binary encoding versus $n$-choose-$k$ encoding.} The left phase gate is controlled on four qubits in a dense binary encoding. The right phase gate is controlled on $k-1=2$ qubits in an $n$-choose-$k$ encoding. Suppose, for example, that $2^5=32$ states were encoded with the dense binary encoding. The $n$-choose-$k$ encoding can encode up to $\binom{n}{k}$ states; therefore at least $n=7$ qubits would be needed to encode the same states with the $n$-choose-$k$ encoding because there exists no $k$ such that $\binom{n}{k} \geq 32$ for $n \leq 6$. For $n=7$, the minimum sufficient value of $k$ is $k=3$, in which case $\binom{7}{3} = 35$. The figure above could represent a re-encoding $\ket{10110} \to \ket{1001001}$.}
    \label{nchoosek}
\end{figure}

To balance the number of qubits necessary\footnote{Facing large qubit overheads for second-quantized approaches is not unique to our problem. Different efficient encoding schemes have been developed to tackle this. See, for example \cite{PhysRevA.104.042607}, or for examples in quantum chemisty see \cite{PhysRevResearch.4.023154, Huang_2023}.} and the subsequent number of quantum gates we use what we call an ``$n$-choose-$k$'' encoding. (See figure~\ref{nchoosek} for a sketched comparison with the dense binary encoding.) Let $F_i$ be the set of all Fock states that have a specific population ``$i$'' of particle species. For example, $F_\gamma$ would be all single-photon Fock states and $F_{e^-e^+}$ would be all electron-positron pair Fock states. We designate $n_i$ qubits for $F_i$. Then we choose $n_i \geq k_i > 0$. States of $F_i$ are then encoded by bitstrings where all $F_{j \neq i}$ qubits are in the $\ket{0}$ state, and $k_i$ of the $n_i$ qubits are in the $\ket{1}$ state. This scheme encodes $\binom{n_i}{k_i}$ $F_i$ states.

For our simulations, we fix $n_\gamma=2$ and choose $k_\gamma=1$. In particular,
\begin{equation}
    \ket{\gamma_k^{(1)}} = \ket{10} \underbrace{\ket{000 \dots 000}}_{F_{e^-e^+} \text{ Qubits}} \qquad \ket{\gamma_k^{(2)}} = \ket{01} \underbrace{\ket{000 \dots 000}}_{F_{e^-e^+} \text{ Qubits}}
\end{equation}
Meanwhile, $n_{e^-e^+}$ and $k_{e^-e^+}$ can vary. This way we translate creation and annihilation operators into a tensor-product of ladder operators:
\begin{equation}
    [c^\dag b a] \sim \ketbra{\gamma}{e^-e^+} \to \sigma^+ \otimes \underbrace{\sigma^- \otimes \cdots \otimes \sigma^-}_{k_{e^-e^+}}
\end{equation}
or projectors:
\begin{equation}
    [a^\dag a], [b^\dag b] \sim \ketbra{e^-e^+}{e^-e^+} \to \underbrace{\pi^1 \otimes \cdots \otimes \pi^1}_{k_{e^-e^+}}
\end{equation}
where $n_i-k_i$ identity operators are implicit. Here, we use
\begin{equation}
    \sigma^+ = \ketbra{1}{0} \qquad \sigma^- = \ketbra{0}{1} \qquad \pi^0 = \ketbra{0}{0} \qquad \pi^1 = \ketbra{1}{1}
\end{equation}
The $n$-choose-$k$ encoding requires the knowledge of Hamiltonian matrix elements; these will be the time-dependent coefficients of the operators shown above. As such, any possible fermionic phase factors are baked into these elements, and so the fermionic anticommutation relations are directly imposed throughout the simulation.

We now have an encoding of our truncated Fock space and know how the Hamiltonian acts on the computational basis states. We move on to construct a quantum circuit for the time evolution operator that acts on the encoded qubits. To perform time evolution as in \eqref{time_evo_trotter}, we need the quantum gates that implement the unitary $e^{-\frac{i}{2}H(x_k^+)\Delta x^+}$. However, this is difficult to do exactly, so we use an additional first-order Trotterization\footnote{For other quantum algorithms used to carry out time evolution in quantum simulations see e.g.~\cite{PhysRevA.111.022419, PRXQuantum.2.040203}.}:
\begin{equation}
    e^{-\frac{i}{2}H(x_k^+)\Delta x^+} = \prod_{h \in H_0(x_k^+)} e^{-\frac{i}{2}h\Delta x^+} \prod_{h \in V(x_k^+)} e^{-\frac{i}{2}h\Delta x^+} \prod_{h \in H^c(x_k^+)} e^{-\frac{i}{2}h\Delta x^+} + \mathcal{O}[(\Delta x^+)^2]
    \label{instant_trot}
\end{equation}
Each $h$ is a Hermitian operator that acts on two Fock states. For instance, $C \ketbra{e^-e^+}{e^-e^+}$ and $C \ketbra{e^-e^+}{\gamma} + C^* \ketbra{\gamma}{e^-e^+}$ are examples of $h$ operators. Now we turn to the problem of finding the quantum gates that implement each $e^{-\frac{i}{2}h\Delta x^+}$.

\begin{figure}[t]
    \centering
    $e^{-\frac{i}{2} C \pi^1 \pi^1 \pi^1 \pi^1} = $
    \begin{quantikz}[align equals at=2.5, row sep={24pt, between origins}]
        & \ctrl{3} & \\
        & \control{} & \\
        & \control{} & \\
        & \gate{P(-\frac{C}{2})} &
    \end{quantikz}
    \caption{\textbf{A quantum gate used for time evolution with the free Hamiltonian.} In this four-qubit example, the gate applies $e^{-\frac{i}{2}C}$ only on the computational basis state $\ket{1111}$ and leaves all other fifteen basis states alone. In our delta-pulse simulations, $C(x^+)$ includes either $P^-+Q^-$ or $p^-+q^-$ depending on whether the background field is nonzero at time $x^+$.}
    \label{H0_mcp}
\end{figure}
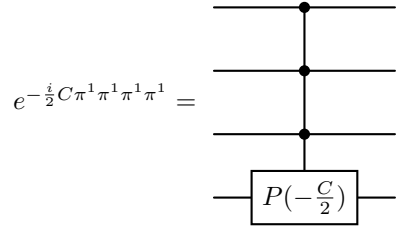

In the free Hamiltonian, the only operators are of the type $C \ketbra{e^-e^+}{e^-e^+}$. The matrix exponential of a string of projectors is a multi-controlled phase gate as shown in figure~\ref{H0_mcp}.
The phase gate $P(\theta) = e^{i\theta \pi^1}$ applies a phase of $e^{i\theta}$ on $\ket{1}$ but leaves $\ket{0}$ alone.

\begin{figure}[h]
    \centering
    \begin{subfigure}[b]{\textwidth}
        \centering
        \resizebox{\textwidth}{!}{
            \begin{quantikz}[align equals at=1.5, row sep={24pt, between origins}]
                & \ctrl{1} & \\
                & \gate{R(a,b)} &
            \end{quantikz}
            $ = $ \begin{quantikz}[align equals at=1.5, row sep={24pt, between origins}]
                & & & \ctrl{1} & & \ctrl{1} & & \\
                & \gate{R_Z(-\atan(b,a) - \frac{\pi}{2})} & \gate{R_Y(-\frac{1}{2}\sqrt{a^2+b^2})} & \targ{} & \gate{R_Y(\frac{1}{2}\sqrt{a^2+b^2})} & \targ{} & \gate{R_Z(\atan(b,a) + \frac{\pi}{2})} &
        \end{quantikz}}
    \end{subfigure}
    
    \vspace{24pt}
    
    \begin{subfigure}[b]{\textwidth}
        \centering
        \begin{quantikz}[align equals at=1.5, row sep={24pt, between origins}]
            & \ctrl{1} & \\
            & \gate{R_X(a)} &
        \end{quantikz}
        $ = $ \begin{quantikz}[align equals at=1.5, row sep={24pt, between origins}]
            & & & \ctrl{1} & & \ctrl{1} & & \\
            & \gate{R_Z(-\frac{\pi}{2})} & \gate{R_Y(-\frac{a}{2})} & \targ{} & \gate{R_Y(\frac{a}{2})} & \targ{} & \gate{R_Z(\frac{\pi}{2})} &
        \end{quantikz}
    \end{subfigure}
    
    \vspace{24pt}
    
    \begin{subfigure}[b]{\textwidth}
        \centering
        \begin{quantikz}[align equals at=1.5, row sep={24pt, between origins}]
            & \ctrl{1} & \\
            & \gate{R_Y(b)} &
        \end{quantikz}
        $ = $ \begin{quantikz}[align equals at=1.5, row sep={24pt, between origins}]
            & & \ctrl{1} & & \ctrl{1} & \\
            & \gate{R_Y(\frac{b}{2})} & \targ{} & \gate{R_Y(-\frac{b}{2})} & \targ{} &
        \end{quantikz}
    \end{subfigure}
    \caption{\textbf{Controlled rotations when the rotation angles are complex ($a+ib$), real, or imaginary.} These decompositions are useful when quantum hardware does not have native support for controlled unitaries. These gates can be extended to multi-controlled gates by using multi-controlled CNOTs.}
    \label{controlled_rotations}
\end{figure}
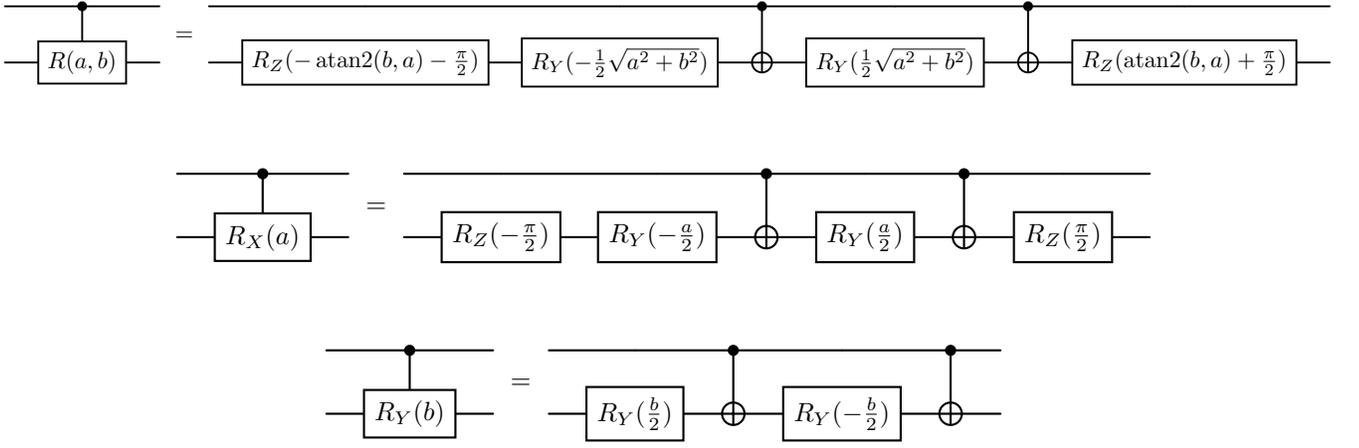

\begin{figure}[h]
    \centering
    $e^{-\frac{i}{2} [ C \sigma^+ \sigma^- \sigma^- \sigma^- + \text{ H.c.} ]} = $
    \begin{quantikz}[align equals at=2.5, row sep={24pt, between origins}]
        & & & \targ{} & \ctrl{3} & \targ{} & & & \\
        & & \targ{} & & \control{} & & \targ{} & & \\
        & \targ{} & & & \control{} & & & \targ{} & \\
        & \ctrl{-1} & \ctrl{-2} & \ctrl{-3} & \gate{R(a,b)} & \ctrl{-3} & \ctrl{-2} & \ctrl{-1} &
    \end{quantikz}
    \caption{\textbf{Givens rotation circuit that performs a portion of the time evolution with the interaction Hamiltonian.} The CNOTs diagonalize the string of ladder operators shown in the exponential. They entangle the qubit to be rotated with the rest of the qubits. Note that $C=a+ib$, and in our quantum simulations, $C$ is a coefficient found in the interaction Hamiltonian \eqref{interaction_hamiltonian}.}
    \label{givens_rotation}
\end{figure}
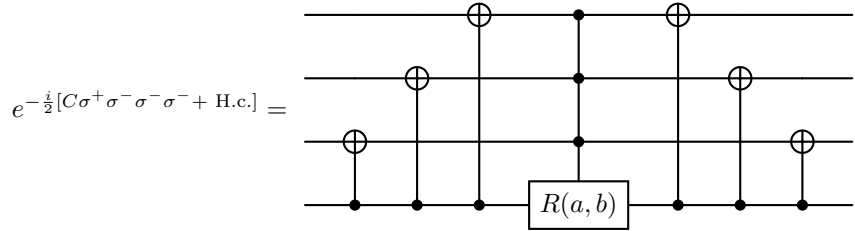

In the interaction Hamiltonian, the operators drive transitions between electron-positron pairs and photons. They appear as $C \ketbra{e^-e^+}{\gamma} + C^* \ketbra{\gamma}{e^-e^+}$, where $C$ is a complex coefficient. The matrix exponential of this operator includes a multi-controlled rotation gate. If $C$ were purely real, then the rotation gate would be an $R_X(\theta) = e^{-i\frac{\theta}{2}X}$ gate. If $C$ were purely imaginary, then the rotation gate would be an $R_Y(\theta) = e^{-i\frac{\theta}{2}Y}$ gate. For a complex coefficient, we define
\begin{equation}
    R(a,b) = e^{-\frac{i}{2}(aX + bY)} = R_Z(\atan(b,a)) R_X(\sqrt{a^2+b^2}) R_Z(-\atan(b,a))
\end{equation}
where $R_Z(\theta) = e^{-i\frac{\theta}{2}Z}$. The controlled versions of these rotation operators are shown in figure~\ref{controlled_rotations}.
However, the matrix exponentials of these Hamiltonian operators are actually Givens rotations \cite{Nielsen_Chuang_2010} shown in figure~\ref{givens_rotation}.
Givens rotation gates rotate two different computational basis states among each other, leaving all other computational basis states alone. They are effectively generalizations of the single-qubit $R_X$ and $R_Y$ gates.

Finally, in the counterterm Hamiltonian, there are $C \ketbra{\gamma}{\gamma}$ and $C\ketbra{\gamma}{\gamma'} + C\ketbra{\gamma'}{\gamma}$ operators with real coefficients (see Sec.~\ref{counterterms}). Matrix exponentials of the former give single-qubit phase gates. Matrix exponentials of the latter lead to a two-qubit Givens rotation gates.

Current quantum hardware do not support multi-controlled CNOTs or phase gates. There exist methods \cite{zindorf2025efficientimplementationmulticontrolledquantum} to decompose them into CNOTs and $T$ gates that scale linearly in the number of control qubits. Moreover, using ancillas has shown to have significant discounts in these decompositions. Thus, the quantum resources necessary in our quantum simulations predominantly depend on the number $N_e$ of electron-positron pairs. Per time step, there are $N_e$ multi-controlled phase gates, $2N_e$ Givens rotations, and the handful of gates from the counter Hamiltonian. For each gate, the number of control qubits is essentially $k_{e^-e^+}$. If $N_t$ is the number of time steps, then the circuit depth and wall time scale with $k_{e^-e^+}N_eN_t$. If $N_e \approx \binom{n_{e^-e^+}}{k_{e^-e^+}}$, then we conclude that the quantum resources scale polynomially in the number of qubits.

\begin{figure}[t!]
    \centering
    \includegraphics[width=\textwidth]{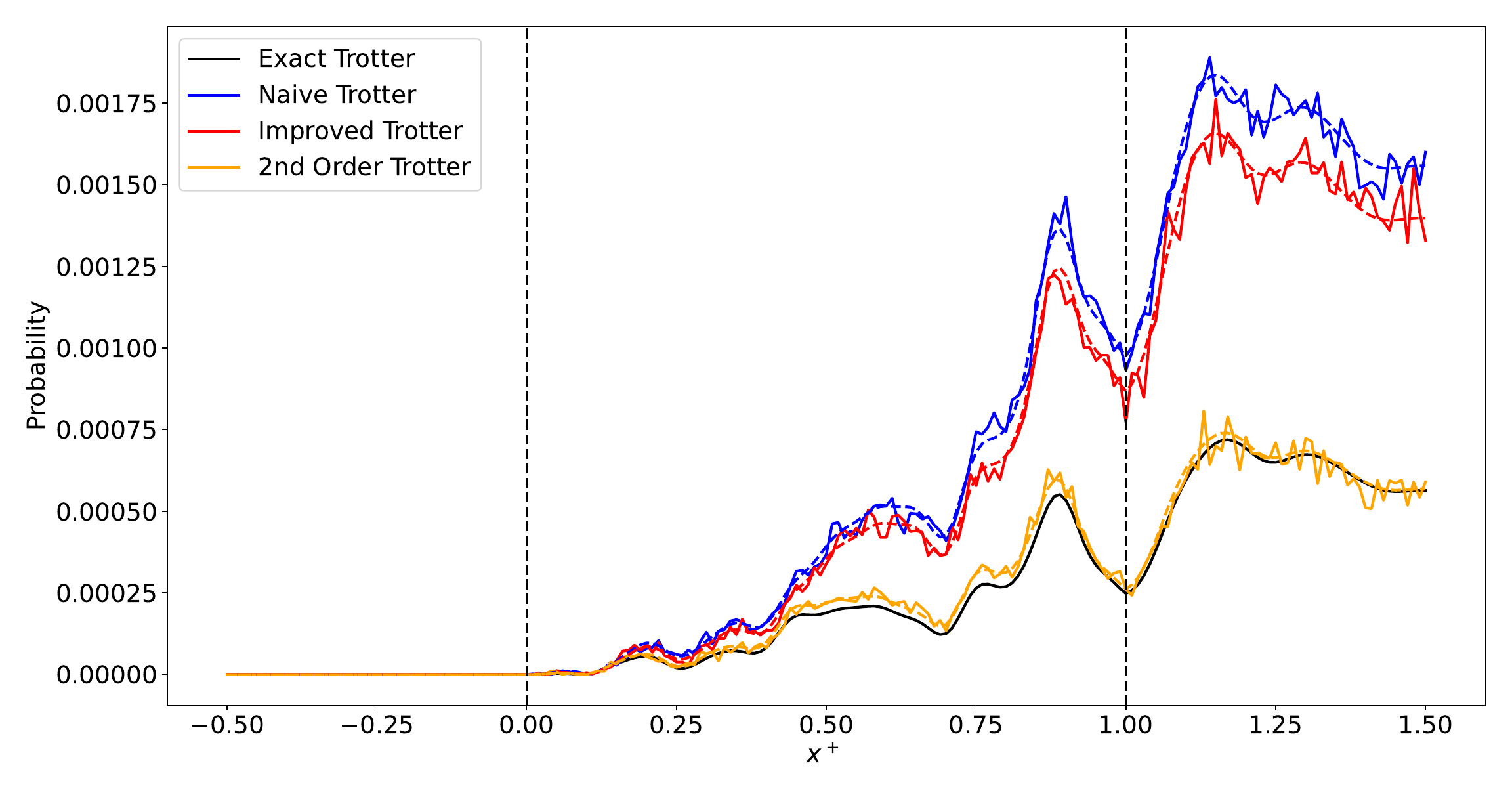}
    \caption{\textbf{The importance of second-order Trotterization.} The figure above shows the time-dependent polarization flip probabilities obtained from (noiseless) quantum and classical simulations. Here $\Delta x^+ = 0.01$ and $T=1$. The coupling was increased to $e=2$ to raise the probabilities and lower the shot requirements. $5 \times 10^5$ shots were used for measurements. The dashed vertical lines show when the first and second pulses occur, as well as the moments when adiabatic turn-on ends and adiabatic turn-off begins. The squared $S$-matrix element (probability) is taken to be the final data point of each curve, once adiabatic turn-off is complete. ``Exact Trotter'' shows probabilities classically computed with \eqref{time_evo_trotter}. ``Naive Trotter'' and ``Improved Trotter'' show two implementations of \eqref{instant_trot} as explained in the text. ``2nd Order Trotter'' shows the implementation of \eqref{second_order}. The dashed curves show what the quantum probabilities would be without shot noise. We see that second order Trotterization is needed for this $\Delta x^+$ in order to converge to the result of exact Hamiltonian exponentiation.}
    \label{quantum_sim}
\end{figure}

In quantum simulations, we are limited by the number of measurements, or shots, we can perform on a quantum circuit. Rare processes like the helicity flip could require tens of millions of measurements per circuit with physical parameters. To bring the cost down, we scale up the coupling to $e=2$, which increases the probabilities by a factor of about $10^3$. Figure~\ref{quantum_sim} shows a noiseless quantum simulation performed with $(K,N)=(3,2)$ and $T=1$. A total of 200 intermediate electron-positron pair states were included in the simulation. With $n_{e^-e^+}=10$ and choosing $k_{e^-e^+}=4$, we performed a 12-qubit simulation. 

The resource costs for a single Trotter-step circuit, $e^{-\frac{i}{2}H(x_k^+)\Delta x^+}$, were as follows. Before transpilation into purely single- and two-qubit gates, the circuit depth was about 3500 gates, of which about 1000 were expressed as multi-controlled unitaries. After a pass through the Qiskit  transpiler~\cite{qiskit2024} (without ancillas), the depth is of the order of 60,000 gates, of which about 26,000 were two-qubit (CNOT) gates and 14,500 were $T$ gates. If two ancilla qubits are allowed and transpilation is done using Qiskit's implementation of the v-chain algorithm \cite{bennakhi2024analyzingquantumcircuitdepth} to decompose the multi-controlled unitary gates, the circuit depth can be  brought down to about 23,500 gates, with about 11,000 CNOT and 5000 $T$ gates.

For completeness, we estimate the quantum resources needed to implement a single Trotter-step circuit in a simulation utilizing the Jordan-Wigner (JW) transformation. Although two qubits are still needed for the photons, the electrons and positrons each need 200 qubits to encode all helicity and momentum modes used in our discretization. Therefore, a total of 402 qubits would have been required, if we had gone the JW route -- this cost is the main reason we opted for the $n$-choose-$k$ encoding. With JW, time evolution generated by the free Hamiltonian can be implemented with a total of 400 (single-qubit) phase gates. The time evolution generated by the interaction Hamiltonian would require 400 Givens rotations (200 momentum-conserving electron-positron pairs interacting with two photons) acting on three qubits. Each Givens rotation can be nominally decomposed into four CNOTs and six rotation gates, assuming all-to-all connectivity. However, each Givens rotation comes attached with a product of C$Z$ gates that account for fermionic statistics. The exact number depends on the ordering of the Fock basis; for example, we could take the ordering in \eqref{fock_state}. We estimate of order $\sim 20$ C$Z$ gates per Givens rotation. Time evolution from the counterterm Hamiltonian once again costs two phase gates and two two-qubit Givens rotations acting on the photon qubits. The total cost is roughly 3000 single-qubit gates and 10,000 two-qubit gates, with the dominant cost likely coming from the C$Z$'s of the JW transformation. Note that in this case, the single- and two-qubit gate decomposition does not include $T$ gates, prior to the synthesis of rotation and phase gates.

The quantum simulation used $\Delta x^+=0.01$ for a total of $N_t=200$ time steps. We find that using \eqref{instant_trot} leads to significant Trotter error at late times. Moreover, changing the ordering of the decomposition of $H(x_k^+)$ into $h$ operators can modify the Trotter error. In figure~\ref{quantum_sim}, we demonstrate this phenomenon by using a ``Naive'' decomposition of $H(x_k^+)$ as well as a better motivated ``Improved'' decomposition. The ``Naive'' decomposition Gray-code \cite{Nielsen_Chuang_2010} orders the Givens rotation gates involved in the interaction Hamiltonian evolution according to the relative CNOTs between them. (This would permit a generally greater chance of CNOT cancellations than without the ordering.) With the ``Improved'' decomposition, we recognize that some elements of $H^2$ may vanish by a sum of a equal and opposite momenta. Once Trotterized, a product such as $h_i h_j$ may substantially lose this structure. To mitigate this, we order our $h$ operators in the improved decomposition such that subsequent $h$ operators have equal and opposite momenta as much as possible.

Nevertheless, the first-order Trotter error is severe. One option is to use a smaller time step; for example, $\Delta x^+ \to 0.001$. However, this could be too costly. For this reason, we implemented the second-order \cite{Ikeda2023minimum} version of \eqref{instant_trot}:
\begin{equation}
    e^{-\frac{i}{2}H(x_k^+) \Delta x^+} = \prod_{h=h_1 \cdots h_N} e^{-\frac{i}{4} h \Delta x^+} \prod_{h=h_N \cdots h_1} e^{-\frac{i}{4} h \Delta x^+} + \mathcal{O}[(\Delta x^+)^3]
    \label{second_order}
\end{equation}
This second-order Trotterization -- along with the aforementioned improved decomposition -- dramatically reduces the Trotter error, even agreeing with the ``Exact'' Trotterization within shot noise. In terms of quantum resources, the second-order product formula adds a factor of two to our estimates above.

\section{Conclusion}
\label{conclusion}

In this paper we have investigated the analytical and computational resources required to perform a Hamiltonian simulation of a one-loop SFQED process. The lightfront Hamiltonian was used to perform an adiabatic time evolution of an asymptotic photon Fock state. The main results of our work can be summarized as follows:
\begin{itemize}
\item Simulating loop-level processes requires confronting symmetry-violating terms stemming from hard momentum cut-offs which are natural and necessary in numerics.  We derived counterterms that cancel these spurious effects across a large class of background fields and implemented them in the simulations.

\item Devising efficient simulations at loop level requires determining  rough boundaries of the space of ``relevant" intermediate states for the process of interest, so that the truncation and map to the Hilbert space of the quantum computer reflects these boundaries. In this work we used perturbative formulae to characterize the relevant subspace of intermediate states for given background field strengths. 

\item Efficient simulation also requires some optimization of both the qubit encoding and the map from the continuum time evolution operator to quantum gates. 
We introduced a new $n$-choose-$k$ encoding to balance qubit count with gate depth, and implemented a time evolution circuit based on multi-controlled rotations which allow for a range of optimizations. 

\item We executed noiseless quantum simulations in order to benchmark truncation and discretization effects and estimate quantum resources required for a full simulation. The number of qubits required for our simulations is practical for near-term quantum devices. However, our circuits are too deep for current quantum computers, and improved gate fidelities and error correction are needed to perform real quantum simulations.
\end{itemize}
We close by commenting on some directions for future work. 

First, while were able to find appropriate counterterms for our simulations, it would be valuable to have more systematic methods to compute the relevant counterterms for higher-order SFQED processes. Ideally such methods would rely less on perturbation theory than we have, especially if the goal is to simulate nonperturbative phenomena or to tackle regimes of extremely high intensity where the Furry expansion is conjectured to break down~\cite{RitusRN1, Narozhnyi:1980dc, fedotov:2017:_ritus_narozhny}.

Other complications with renormalization can arise with sufficiently aggressive truncations of Fock space, which eliminates some radiative corrections at a given order while retaining others~\cite{Perry:1990mz} (see~\cite{Hiller:2016itl} for a review of the issues and resolutions.) Fortunately, these complications did not arise for one-loop vacuum polarization in the light-front. However, we also note that quantum simulations based on occupation number registers -- independent registers of qubits encoding occupation numbers for each single-particle Fock state -- probably avoid these issues in a practical sense: to high loop order, no spurious diagram elimination arises.

Quantum computers may one day aid in computing observables in complex systems such as cascades~\cite{Bell:2008zzb, PhysRevSTAB.14.054401, Mironov:2021fft, PhysRevE.110.065208, PhysRevX.15.011062, PhysRevLett.134.135001}. Quantum simulation of these processes would require many more qubits, stemming from the high dimensionality of the multiparticle Hilbert space that the processes explore. It would be interesting to assess the quantum resources needed for an accurate cascade simulation. In particular, it would be useful to compare different qubit encodings for such processes, investigating, for example, the feasibility of  Jordan-Wigner versus other encodings.

Relatedly, a comparison of the challenges facing the momentum-space approach with those found in  position-space lattice formulations~\cite{PhysRevD.11.395, PhysRevD.91.054506, Magnifico_2021} should be carried out. Rapid progress is being made on simulating spatial lattice gauge theories on quantum computers \cite{l8b6-h5s5, davoudi2025quantumcomputationhadronscattering, PhysRevD.100.034518, 1km8-3tc3, PhysRevD.109.114510, PhysRevD.110.034515, ciavarella2025efficienttruncationssunclattice, PhysRevD.111.074516, PhysRevD.103.094501, jiang2025nonabeliandynamicscubeimproving, PhysRevD.104.034504, jiang2025nonabeliandynamicscubeimproving, balaji2025quantumcircuitssu3lattice, PhysRevD.111.114516, li2024sequencyhierarchytruncationseqht, crippa2024analysisconfinementstring2, gustafson2024improvedfermionhamiltoniansquantum}, and SFQED should also be studied in this framework. Which approach will be most viable for future quantum hardware remains to be seen.

\section*{Acknowledgments}
A.I.~thanks Ben King for useful discussions. This work was supported by the U.S. Department of Energy, Office of Science, Office of High Energy Physics Quantum Information Science Enabled Discovery (QuantISED) program. A.I.~is supported by the STFC consolidated grant ``Particle Theory at the Higgs Centre'' ST/X000494/1. This work used the Delta system at the National Center for Supercomputing Applications through allocation PHY230137 from the Advanced Cyberinfrastructure Coordination Ecosystem: Services \& Support (ACCESS) program, which is supported by National Science Foundation grants \#2138259, \#2138286, \#2138307, \#2137603, and \#2138296.

\bibliographystyle{utphys}
\bibliography{bib}

\appendix

\newpage

\section{Hamiltonian Field Theory}
\label{app:hamiltonian_field_theory}

This appendix demonstrates how some operators, if generated by cutoff effects in an otherwise well-behaved Hamiltonian, can spoil the  symmetry structure and must be cancelled by counterterms. 

In the text, we calculate amplitudes that only involve photon operators (see Sec.~\ref{old_fashioned_perturbation_theory}). Therefore we focus on how the hard momentum cutoffs affect the free Maxwell theory in the light front. Because the cutoffs are not Lorentz covariant, we expect them to generate Lorentz-violating effective operators. We will see that the operators also violate gauge symmetry.
	
Recall the Maxwell Lagrangian density
\begin{equation}
    \mathcal{L} = -\frac{1}{4} F_{\mu\nu} F^{\mu\nu} \;,
\end{equation}
with canonical momentum
\begin{equation}
    \Pi^\mu = \frac{\partial\mathcal{L}}{\partial(\partial_+ A_\mu)} = F^{\mu+} \;.
\end{equation}
The Hamiltonian is
\begin{IEEEeqnarray*}{rCl}
    H & = & \frac{1}{2} \int d^3\mathsf{x} \left( \Pi^\mu \partial_+ A_\mu - \mathcal{L} \right) = \frac{1}{4} \int d^3\mathsf{x} \left( \frac{1}{4} (\Pi^-)^2 + F_{12}^2 - A^- ( \partial_- \Pi^- + \partial_i \Pi^i ) \right)  \;, \yesnumber
\end{IEEEeqnarray*}
We proceed with canonical quantization, imposing the commutation relations
\begin{equation}
    [A_\mu(\mathsf{x}), \Pi^\nu(\mathsf{y})] = i \delta_\mu^\nu \delta^3(\mathsf{x} - \mathsf{y}) \qquad [A^\mu(\mathsf{x}), A^\nu(\mathsf{y})] = 0 = [\Pi^\mu(\mathsf{x}), \Pi^\nu(\mathsf{y})] \;;.
\end{equation}
Our quantized theory has primary constraint $\Pi^+ = 0$. The secondary constraint is
\begin{equation}
    [H, \Pi^+] = 0 \implies \partial_- \Pi^- + \partial_i \Pi^i \equiv G = 0
\end{equation}
which is Gauss' law. There are no further constraints, because $[H,G] = 0$ and $[\Pi^+, G] = 0$. 

\subsection{Counterterms}
	
A covariant regulator would only generate Lorentz- and gauge-invariant effective operators such as $F^{2n}$. Such operators give correspond to higher-derivative interactions in the Maxwell Hamiltonian. Although our amplitudes contain contributions from these operators, they do not account for the ``unphysical contributions'' found in Sec.~\ref{old_fashioned_perturbation_theory}. Instead, we examine the low-dimension effective operators not of the standard form; namely, the  $AA$ and $AA\mathcal{AA}$ operators found in the text.

Suppose we have an effective Hamiltonian of the form
\begin{IEEEeqnarray*}{rCl}
    H & = & \frac{1}{4} \int d^3\mathsf{x} \ \bigg( \frac{1}{4} (\Pi^-)^2 + F_{12}^2 - A^- ( \partial_- \Pi^- + \partial_i \Pi^i ) \yesnumber \\
    & & \qquad \qquad + \ C_{AA} A_\mu A^\mu + C_{AA\mathcal{AA}} (A_\mu A^\mu) (\mathcal{A}_\nu \mathcal{A}^\nu) + C_{2A\mathcal{A}} (A_\mu \mathcal{A}^\mu)^2 + D_{\mu\nu\rho\sigma} A^\mu A^\nu \mathcal{A}^\rho \mathcal{A}^\sigma \bigg)
\end{IEEEeqnarray*}
where $D_{\mu\nu\rho\sigma}$ is symmetric and purely off-diagonal in the $\mu,\nu$ indices, and likewise in the $\rho,\sigma$ indices. Furthermore, $(+,-)$ indices do not mix with $(1,2)$ indices. The secondary constraint from demanding $[H, \Pi^+] = 0$ is now
\begin{equation}
    \partial_- \Pi^- + \partial_i \Pi^i - C_{AA} A^+ - C_{AA\mathcal{AA}} \mathcal{A}_\mu \mathcal{A}^\mu A^+ - C_{2A\mathcal{A}} A_\mu \mathcal{A}^\mu \mathcal{A}^+ - 2D_{+-\rho\sigma} A^+ \mathcal{A}^\rho \mathcal{A}^\sigma \equiv G = 0.
\end{equation}
However, $G$ does not commute with $H$:
\begin{IEEEeqnarray*}{rCl}
    -2i[H, G] & = & C_{AA} \left( 2\partial_- A^- + \partial_i A^i + \frac{1}{4} \Pi^- \right) \yesnumber \\
    & + & C_{AA\mathcal{AA}} \left( \partial_-(\mathcal{A}_\mu \mathcal{A}^\mu A^-) + \partial_i (\mathcal{A}_\mu \mathcal{A}^\mu A^i) + \frac{1}{4} \mathcal{A}_\mu \mathcal{A}^\mu \Pi^- + \mathcal{A}_\mu \mathcal{A}^\mu \partial_- A^- \right) \\
    & + & C_{2A\mathcal{A}} \left( \partial_-(A_\mu \mathcal{A}^\mu \mathcal{A}^-) + \partial_i(A_\mu \mathcal{A}^\mu \mathcal{A}^i) + \frac{1}{8} \mathcal{A}^+ \mathcal{A}^- \Pi^- + \frac{1}{2} \mathcal{A}^+ \mathcal{A}^- \partial_- A^- + \frac{1}{2} \mathcal{A}^+ \mathcal{A}^- \partial_i A^- \right) \\
    & + & 2D_{+-\rho\sigma} \partial_-(A^- \mathcal{A}^\rho \mathcal{A}^\sigma) - D_{ij\rho\sigma} \partial_i(A^j \mathcal{A}^\rho \mathcal{A}^\sigma) -D_{+-\rho\sigma} \Pi^- \mathcal{A}^\rho \mathcal{A}^\sigma - 2D_{+-\rho\sigma} \partial_- A^- \mathcal{A}^\rho \mathcal{A}^\sigma.
\end{IEEEeqnarray*}
Even in the light-front gauge, this commutator does not vanish. This introduces more constraints; however, each constraint removes a degree of freedom of the photon field. $\Pi^+ = 0$ and $G = 0$ leave the ordinary Maxwell theory with two photon polarizations. In the effective theory, this is no longer the case. Therefore, the effective operators do not preserve the gauge structure of the theory. To fix this, we adjust counterterms for the effective operators, matching the coefficients to the relevant parts of the amplitudes (see Sec.~\ref{counterterms}) in order to cancel off the gauge violating effects of the momentum cutoffs.

\subsection{Spacetime Symmetry Breaking}

Let $K_1,K_2,K_3$ and $J_1,J_2,J_3$ be generators of Lorentz boosts and rotations, respectively. Then generators of the Lorentz group can be defined as \cite{kogut:1970:infinite_momentum}
\begin{IEEEeqnarray*}{CCCCC}
    S_1 = \frac{K_1 - J_2}{\sqrt{2}} & \qquad & S_2 = \frac{K_2 + J_1}{\sqrt{2}} & \qquad & K_3 \\
    B_1 = \frac{K_1 + J_2}{\sqrt{2}} & \qquad & B_2 = \frac{K_2 - J_1}{\sqrt{2}} & \qquad & J_3 \yesnumber
\end{IEEEeqnarray*}
The Maxwell Hamiltonian obeys the following commutation relations:
\begin{equation}
    [K_3, H] = iH \qquad [S_k, H] = 0 \qquad [B_k, H] = -iP^k \qquad [J_3, H] = 0
    \label{lorentz_commutators}
\end{equation}
where $P^k$ is the generator of (transverse) translations. After fixing the light-front gauge, $A^+ = 0 = \mathcal{A}^+$, as well as using Gauss' law, $\partial_-^2 A^- + \partial_- \partial_i A^i = 0$, the effective operators contain operators of the form
\begin{IEEEeqnarray*}{rCl}
    X & \equiv & \int d^3\mathsf{x} \ A^1A^2 + A^2A^1 \qquad (\text{flip}) \\
    O & \equiv & \int d^3\mathsf{x} \ A^1A^1 + A^2A^2 \qquad (\text{no flip}) \yesnumber
\end{IEEEeqnarray*}
We will now check whether these operators respect \eqref{lorentz_commutators}; if they do not, then they break spacetime symmetries.

\subsubsection*{$K_3$ Transformations}

From the equations of motion, $\partial_\mu F^{\mu i} = 0$, we have
\begin{equation}
    4\partial_- \partial_+ A_i = \partial_i \partial_- A^- + \partial_j F_{ji} \implies \partial_+ A_i = \frac{1}{4} \partial_\perp^2 \Lambda_i
\end{equation}
where $\Lambda_i$ satisfies $\partial_- \Lambda_i = A_i$. Then when integrating by parts,
\begin{IEEEeqnarray*}{rCl}
    \partial_+ X & = & \frac{1}{4} \int d^3\mathsf{x} \ \bigg( \partial_-( \Lambda_1 \partial_\perp^2 \Lambda_2 ) + \partial_-( \Lambda_2 \partial_\perp^2 \Lambda_1 ) \bigg) = 0 \\ [12pt]
    \partial_+ O & = & \frac{1}{4} \int d^3\mathsf{x} \ \bigg( \partial_-( \Lambda_i \partial_\perp^2 \Lambda_i ) \bigg) = 0  \yesnumber
\end{IEEEeqnarray*}
Therefore, $X$ and $O$ are conserved.

Now using
\begin{equation}
    i[K_3, A^i] = (x^- \partial_- - x^+ \partial_+) A^i
\end{equation}
we find that
\begin{IEEEeqnarray*}{rCl}
    i[K_3, X] & = & -X + \int d^3\mathsf{x} \ \bigg( \partial_-(x^-A^1A^2 + x^-A^2A^1) - x^+ \partial_+(A^1A^2 + A^2A^1) \bigg) \\
    & = & -X \\ [12pt]
    i[K_3, O] & = & -O + \int d^3\mathsf{x} \ \bigg( \partial_-(x^-A^1A^1 + x^-A^2A^2) - x^+ \partial_+(A^1A^1 + A^2A^2) \bigg) \\
    & = & -O \yesnumber
\end{IEEEeqnarray*}
by boundary conditions.

\subsubsection*{$S_k$ Transformations}

Using
\begin{IEEEeqnarray*}{rCl}
    -i[S_k, A^i] & = & (x^k \partial_+ + x^- \partial_k) A^i - \delta_{ik} A^- + \partial_i \Lambda_k \\
    & = & (x^k \partial_+ + x^- \partial_k) A^i - \delta_{ik} \partial_j \Lambda_j + \partial_i \Lambda_k \yesnumber
\end{IEEEeqnarray*}
we find that
\begin{IEEEeqnarray*}{rCl}
    -i[S_k, X] & = & \int d^3\mathsf{x} \ \bigg( x^k \partial_+(A^1A^2 + A^2A^1) + x^- \partial_k(A^1A^2 + A^2A^1) \\
    & + & \delta_{1k}(\partial_i \Lambda_i \partial_- \Lambda_2 + \partial_- \Lambda_2 \partial_i \Lambda_i) + \delta_{2k}(\partial_i \Lambda_i \partial_- \Lambda_1 + \partial_- \Lambda_1 \partial_i \Lambda_i) \\
    & - & \partial_1 \Lambda_k \partial_- \Lambda_2 - \partial_2 \Lambda_k \partial_- \Lambda_1 - \partial_- \Lambda_1 \partial_2 \Lambda_k - \partial_- \Lambda_2 \partial_1 \Lambda_k \bigg) \\
    & = & \int d^3\mathsf{x} \ \bigg( x^k \partial_+(A^1A^2 + A^2A^1) + f(\partial\Lambda) \bigg) \\ [12pt]
    -i[S_k, O] & = & \int d^3\mathsf{x} \ \bigg( x^k \partial_+(A^1A^1 + A^2A^2) + x^- \partial_k(A^1A^1 + A^2A^2) \\
    & + & \delta_{1k}(\partial_i \Lambda_i \partial_- \Lambda_1 + \partial_- \Lambda_1 \partial_i \Lambda_i) + \delta_{2k}(\partial_i \Lambda_i \partial_- \Lambda_2 + \partial_- \Lambda_2 \partial_i \Lambda_i) \\
    & - & \partial_i \Lambda_i \partial_- \Lambda_k - \partial_- \Lambda_k \partial_i \Lambda_i \bigg) \\
    & = & \int d^3\mathsf{x} \ \bigg( x^k \partial_+(A^1A^1 + A^2A^2) \bigg) \yesnumber
\end{IEEEeqnarray*}
by boundary conditions. The $x^k$ in the integrand prohibits the light-front time derivatives from vanishing.

\subsubsection*{$B_k$ Transformations}

Using
\begin{equation}
    i[B_k, A^i] = (x^+ \partial_k + x^k \partial_-) A^i
\end{equation}
we find that
\begin{IEEEeqnarray*}{rCl}
    i[B_k, X] & = & \int d^3\mathsf{x} \ \bigg( x^+ \partial_k(A^1A^2 + A^2A^1) + x^k \partial_-(A^1A^2 + A^2A^1) \bigg) = 0 \\ [12pt]
    i[B_k, O] & = & \int d^3\mathsf{x} \ \bigg( x^+ \partial_k(A^1A^1 + A^2A^2) + x^k \partial_-(A^1A^1 + A^2A^2) \bigg) = 0 \yesnumber
\end{IEEEeqnarray*}
by boundary conditions.

\subsubsection*{$J_3$ Transformations}

Using
\begin{equation}
    i[J_3, A^i] = (x^2 \partial_1 - x^1 \partial_2) A^i - \epsilon_{ij} A^j
\end{equation}
we find that
\begin{IEEEeqnarray*}{rCl}
    i[J_3, X] & = & \int d^3\mathsf{x} \ \Big( x^2 \partial_1(A^1A^2 + A^2A^1) - x^1 \partial_2(A^1A^2 + A^2A^1)
    - \ \epsilon_{1i}(A^iA^2 + A^2A^i) - \epsilon_{2i}(A^iA^1 + A^1A^i) \Big)\\
    & = & 2\int d^3\mathsf{x} \ A^1A^1 - A^2A^2  \,,  \\ [12pt]
    i[J_3, O] & = & \int d^3\mathsf{x} \ \Big( x^2 \partial_1(A^1A^1 + A^2A^2) - x^1 \partial_2(A^1A^1 + A^2A^2)
    - \ \epsilon_{1i}(A^iA^1 + A^1A^i) - \epsilon_{2i}(A^iA^2 + A^2A^i) \Big) \\
    & = & 0  \,, \yesnumber
\end{IEEEeqnarray*}
by boundary conditions.

\subsubsection*{Verdict}

We have found that $X$-type and $O$-type operators obey the same commutation relations with $K_3$ as the Maxwell Hamiltonian does. However, they do not commute with $S_k$. Therefore, they break transverse spacetime symmetries. $X$ and $O$ commute with $B_k$, which means they do not contribute to the photon momentum operators. Lastly, $X$ does not commute with $J_3$ and thus additionally breaks rotational symmetry. Thus the effective operators that contribute to polarization flip amplitudes break spacetime symmetries that should be preserved, and we again conclude that they represent unphysical effects of the cutoff.

\end{document}